%% file: main.tex
\documentclass[sigplan,screen]{acmart}

\input{snippets/packages-setup}

\author{Fangzheng Lin}
\email{lin.f.f849@m.isct.ac.jp}
\affiliation{
  \institution{Institute of Science Tokyo}
  \state{Tokyo}
  \country{Japan}
}

\author{Zhongfa Wang}
\email{wang.z.e488@m.isct.ac.jp}
\affiliation{
  \institution{Institute of Science Tokyo}
  \state{Tokyo}
  \country{Japan}
}

\author{Hiroshi Sasaki}
\email{sasaki@ict.eng.isct.ac.jp}
\affiliation{
  \institution{Institute of Science Tokyo}
  \state{Tokyo}
  \country{Japan}
}

\copyrightyear{2025}
\acmYear{2025}
\setcopyright{cc}
\setcctype{by}
\acmConference[CGO '25]{Proceedings of the 23rd ACM/IEEE International Symposium on Code Generation and Optimization}{March 01--05, 2025}{Las Vegas, NV, USA}
\acmBooktitle{Proceedings of the 23rd ACM/IEEE International Symposium on Code Generation and Optimization (CGO '25), March 01--05, 2025, Las Vegas, NV, USA}
\acmDOI{10.1145/3696443.3708936}
\acmISBN{979-8-4007-1275-3/25/03}

\ccsdesc[500]{Security and privacy~Vulnerability scanners}
\ccsdesc[300]{Security and privacy~Side-channel analysis and countermeasures}
\ccsdesc[300]{Software and its engineering~Translator writing systems and compiler generators}

\keywords{Spectre gadget detector, binary instrumentation}

\begin{document}

\title[Teapot: Efficiently Uncovering Spectre Gadgets in COTS Binaries]{Teapot: Efficiently Uncovering Spectre Gadgets\\in COTS Binaries}

\input{abstract}

\maketitle

\input{introduction}
\input{background}
\input{threat-model}
\input{motivation}

\input{overview}

\input{two-copy-approach}

\input{tool-arch}
\input{experiments}

\input{limitations}

\input{related-work}
\input{conclusion}

\input{acknowledgements}

\appendix
\input{apdx-case-study}
\input{apdx-adae}

\balance{
\bibliographystyle{ACM-Reference-Format}
\bibliography{main}
}

\end{document}

%% file: snippets/packages-setup.tex
\usepackage{appendix}
\usepackage{balance}
\usepackage{url}
\usepackage{tikz}
\usepackage{pgfplots}
\usepackage{listings}
\usepackage{xcolor}
\usepackage[most]{tcolorbox}
\usepackage{tabularx}
\usepackage{multirow}
\setcitestyle{square,numbers,sort}
\usepackage[nameinlink]{cleveref} %
\usepackage[labelformat=simple,labelfont=bf]{subcaption} %
\usepackage{colortbl}
\usepackage{enumitem}
\usepackage{amsthm}
\usepackage[binary-units]{siunitx}
\usepackage{booktabs}

\theoremstyle{definition}

\usetikzlibrary{arrows, arrows.meta, tikzmark, ext.paths.ortho, matrix, positioning, shapes.geometric, calc, decorations.pathreplacing, calligraphy}
\usepgfplotslibrary{groupplots}

\input{snippets/color-defs}

\input{snippets/lstlisting-defs}
\input{snippets/lstlinebgrd-workaround}

\newcommand{\toolname}{Teapot}
\newcommand{\twocopyname}{Speculation Shadows}
\newcommand{\normalfnname}{Real Copy}
\newcommand{\specfnname}{Shadow Copy}

\newcommand{\eg}{e.g.,~}

\crefname{section}{Section}{Sections}
\crefname{equation}{Equation}{Equations}
\crefname{figure}{Figure}{Figures}
\crefname{table}{Table}{Tables}
\crefname{listing}{Listing}{Listings}
\crefname{appendix}{Appendix}{Appendices}

\renewcommand{\cite}[1]{~[\citenum{#1}]}

\definecolor{gray90}{gray}{0.9} %
\setlist[itemize]{leftmargin=10pt,itemindent=0pt,topsep=2pt,partopsep=2.5pt,parsep=1pt,itemsep=1.5pt,listparindent=\parindent{}}
\setlist[enumerate]{leftmargin=16pt,itemindent=0pt,topsep=2pt,partopsep=2.5pt,parsep=1pt,itemsep=1.5pt,listparindent=\parindent{}}

\newcommand{\nocolorref}[1]{\hypersetup{hidelinks}{\ref*{#1}}}

\pgfplotsset{compat=1.18}

%% file: snippets/color-defs.tex
\definecolor{alizarin}{rgb}{0.82, 0.1, 0.26}
\definecolor{dartmouthgreen}{rgb}{0.05, 0.5, 0.06}

\definecolor{myyellow}{RGB}{238,220,150}
\definecolor{myverylightblue}{RGB}{198,231,255}
\definecolor{mylightblue}{RGB}{123,197,227}
\definecolor{mydarkblue}{RGB}{43,79,125}
\definecolor{mygreen}{RGB}{99,121,81}

%% file: snippets/lstlisting-defs.tex
\newcommand{\lstcolourline}[1]{\rlap{\color{#1}\rule[-.3\baselineskip]{\linewidth}{\baselineskip}}}

\definecolor{codegray}{rgb}{0.5,0.5,0.5}
\definecolor{backcolour}{rgb}{0.95,0.95,0.92}

\lstdefinestyle{my-lst-style-fullwidth}{
    basicstyle=\linespread{0.85}\footnotesize\ttfamily,
    breaklines=true,
    captionpos=b,
    language=c,
    escapechar=@
}

\lstdefinestyle{my-lst-style}{
    style=my-lst-style-fullwidth,
    linewidth=0.47\textwidth
}

\lstdefinestyle{my-lst-style-bg}{
    style=my-lst-style,
    frame=tlbr, framesep=0.1cm, framerule=0pt,
    backgroundcolor=\color{backcolour}
}

\lstdefinestyle{my-lst-style-linenum}{
    style=my-lst-style-bg,
    numberstyle=\tiny\color{codegray},
    numbers=left,
    xleftmargin=18pt
}

\lstdefinelanguage{none}{
  identifierstyle=
}

\lstdefinelanguage
   [x64]{Assembler}     %
   [x86masm]{Assembler} %
   {morekeywords={CDQE,CQO,CMPSQ,CMPXCHG16B,JRCXZ,LODSQ,MOVSXD, %
                  POPFQ,PUSHFQ,SCASQ,STOSQ,IRETQ,RDTSCP,SWAPGS, %
                  xmm0, xmmword, movups, movaps,
                  rax,rdx,rcx,rbx,rsi,rdi,rsp,rbp, %
                  r8,r8d,r8w,r8b,r9,r9d,r9w,r9b, %
                  r10,r10d,r10w,r10b,r11,r11d,r11w,r11b, %
                  r12,r12d,r12w,r12b,r13,r13d,r13w,r13b, %
                  r14,r14d,r14w,r14b,r15,r15d,r15w,r15b,.quad}} %

%% file: abstract.tex
\begin{abstract}
  Speculative execution is crucial in enhancing modern processor performance but can introduce Spectre-type vulnerabilities that may leak sensitive information.
  Detecting Spectre gadgets from programs has been a research focus to enhance the analysis and understanding of Spectre attacks.
  However, one of the problems of existing approaches is that they rely on the presence of source code (or are impractical in terms of run-time performance and gadget detection ability).

  This paper presents \toolname{}, the first Spectre gadget scanner that works on COTS binaries with comparable performance to compiler-based alternatives.
  As its core principle, we introduce \twocopyname{}, a novel approach that separates the binary code for normal execution and speculation simulation in order to improve run-time efficiency. %

  \toolname{} is based on static binary rewriting.
  It instruments the program to simulate the effects of speculative execution and also adds integrity checks to detect Spectre gadgets at run time.
  By leveraging fuzzing, \toolname{} succeeds in efficiently detecting Spectre gadgets.
  Evaluations show that \toolname{} outperforms both performance (more than 20$\times$ performant) and gadget detection ability than a previously proposed binary-based approach.
\end{abstract}

%% file: introduction.tex
\section{Introduction}

Transient execution attacks such as Spectre\cite{SpectrePHTBTB} have exposed a new threat to system security.
Secrets can even be leaked from functionally correct programs by abusing speculative execution, which is one of the powerhouses driving today's microprocessor performance.
Specifically, an attacker could carefully craft branch misprediction(s) to exert code snippets which are bug-free in normal execution, to speculatively run into unintended control flows and
read secret information.
Code snippets potentially vulnerable to such attacks are called Spectre gadgets.

Identifying Spectre gadgets in programs has been a focus of many researchers\cite{SpecFuzz,SpecTaint,Kasper,oo7,Spectector,Cats}.
Much effort has been put towards detecting them by assuming the source code is accessible.
State-of-the-art \textit{dynamic} Spectre gadget detectors that \textit{utilize the source code} include SpecFuzz\cite{SpecFuzz} and Kasper\cite{Kasper}; they rely on the compiler to statically instrument the program in order to simulate the consequences of speculative execution (we call it speculation simulation for short) and \textit{dynamically} detect gadgets at run time.
They expand detection coverage with the help of fuzzers.
Such tools are helpful when the source code is present, but it is not always the case.\looseness=-1

Tools that find gadgets from binaries become paramount when analyzing closed-source security-critical applications, such as banking and cryptocurrency apps, password managers, and device drivers.
They allow researchers to improve the analysis and their understanding of Spectre gadgets in the wild.
Not only when dealing with closed-source programs is binary analysis valuable: it allows inspecting the actual deployed application; compilers and toolchains are not necessarily bug-free, and thus properties proven by static source code analysis might not hold for the compiled version\cite{angr}.\looseness=-1

Unfortunately, the only work that operates at the binary level as of this writing is SpecTaint\cite{SpecTaint}, which lacks both run-time efficiency and program-level information, since it is based on a full-system emulator.
In order to advance research in this area, we set our goal to develop a practical binary-based dynamic Spectre gadget detector that works on commercial off-the-shelf (COTS) binaries without the presence of source code.
To this end, we present \toolname{},%
which is based on static binary rewriting, designed with efficiency as its guiding principle.

Teapot's efficiency can be attributed to a key insight derived from the following observation.
Existing approaches have a common problem: normal execution and speculation simulation that require totally different instrumentations coexist inside a single instance of the final instrumented code, and thus every single instrumentation is guarded with an if condition.
This adds significant performance overhead since the program needs to execute a huge number of branch instructions at run time.

Our key insight is that we can eliminate almost all the guard conditions and greatly accelerate run-time performance by separating the binary code executed in normal execution and speculation simulation.
We call this design principle \twocopyname{}.
We present the detailed design and evaluate the performance of \toolname{}.
It significantly outperforms SpecTaint in both execution efficiency and gadget detection ability; in addition, despite being a binary-only approach, \toolname{} yields results comparable to compiler-based instrumentations, proving its practical viability.

Our contributions include the following:
\begin{itemize}
  \item We propose \twocopyname{}, the framework and design principle behind \toolname{}.
  It is engineered to support static binary instrumentation for detecting Spectre gadgets, by completely separating normal execution and speculation simulation code.
  \item We utilize \twocopyname{} to implement and open source \toolname{}, the first Spectre-V1 gadget detector based on static binary rewriting, designed to be highly efficient.
  \item We evaluate \toolname{} and show that it outperforms the state-of-the-art binary-based approach SpecTaint in run-time efficiency (more than 20$\times$ performant) and detection ability of Spectre gadgets in COTS binaries.
  \footnote{We also analyze and discuss gadgets found in \toolname{} but missed in prior works as case studies to demonstrate the detection superiority of our approach in \cref{sec:apdx-case-studies}.}
\end{itemize}

%% file: background.tex
\section{Background}

\subsection{Spectre attacks}\label{ssec:tea}

\begin{lstlisting}[style=my-lst-style-linenum, caption=A potential Spectre-V1 gadget., float=tp, label={lst:gadget}]
int index = user_input();
if (index < 10) {      // B1: Mispredicted
  secret = foo[index]; // L1: Load Secret
  baz = bar[secret];   // L2: Transmit Secret
}
\end{lstlisting}

Modern out-of-order microprocessors rely on many forms of speculation when executing programs to enhance performance.
For instance, a processor may encounter a branch instruction whose direction depends on preceding instructions yet to be completed.
In such case, it consults the branch predictor to predict the correct direction (and address) and \textit{speculatively} continues execution along the predicted path.
If the prediction turns out to be correct, the processor gains performance over waiting for the branch direction to be resolved.
Otherwise, it discards all the results that are executed speculatively (or \textit{transiently}) as if they have not been executed, and resumes along the correct path.

Spectre attacks leverage such transient executions to leak secret information from programs.
The vulnerability lies in the implementation of microprocessors when discarding the results of transiently executed instructions.
Although they successfully remove architectural changes, traces might be left in the microarchitecture resources such as the caches (which might be later inferred via timing covert/side channels).
The attacker uses this vulnerability to steal sensitive data from a victim program by letting the processor transiently access the secret.

We use the canonical Spectre-V1 (aka bounds check bypass) gadget\cite{SpectrePHTBTB} shown in \cref{lst:gadget} to demonstrate how the attack could hypothetically work.
First, the attacker feeds \texttt{index} with values smaller than 10 to train the branch predictor to make a prediction that the if condition (bounds check; B1) holds.
Next, it supplies a malicious \texttt{index} value greater than 10 to induce a branch misprediction that transiently allows bypassing the bounds check to access a secret value into a register (L1).
The secret value is further used as an index of the next memory access, encoding the value as a location inside the cache (L2).
Even after transiently executed instructions are architecturally wiped out, the attacker can learn the secret value by performing cache attacks such as the Prime+Probe attack\cite{EVICT+TIME_PRIME+PROBE}.

\subsection{Dynamic detection of Spectre gadgets}

\input{figs/lst-specfuzz}

While all conditional branches are potential victims of \mbox{Spectre-V1}, they can be exploited to leak secret information only if the incorrect execution path meets certain criteria (such as the one shown in \cref{lst:gadget}).
Finding such gadgets in real-world programs is not trivial as the criteria are complex.
Static gadget scanning works that rely on matching program source with common Spectre code patterns\cite{oo7,MSVC} can be ineffective, as they introduce false positives and false negatives due to complex variations of Spectre gadgets beyond the description of fixed patterns.
Formal methods and symbolic execution\cite{Spectector,SpecuSym,KLEESpectre,Cats} ensure high detection coverage, but do not scale well with large real-world programs because of the large search space and high computational complexity\cite{SpecFuzz}. \looseness=-1

As a result, dynamic fuzzing-based detection tools have evolved to be state-of-the-art in detecting such gadgets because they achieve a good balance between gadget detection ability and practicality.
They instrument the program to simulate the effects of branch misprediction, add integrity checks to flag Spectre gadgets, and provide random inputs to test the program (with fuzzing).
At a high level they share a similar software architecture shown in \cref{lst:specfuzz}.
It stores the current program state (makes a checkpoint) before each conditional branch.
To simulate branch misprediction, it forces the program onto the reverted branch direction.
Then, speculative execution is simulated by letting the program proceed on the wrong path for some instructions until a rollback point is reached.
The program state from the checkpoint is restored, and normal execution resumes.
We introduce the three major works in this category in the following:

\subsubsection{SpecFuzz}
SpecFuzz\cite{SpecFuzz} is the first to propose dynamically simulating the effects of branch misprediction in software to detect Spectre gadgets.
It does so by adding instrumentation to the analyzed programs during compilation (and thus requires source code).
During speculation simulation, SpecFuzz detects out-of-bounds memory errors with AddressSanitizer~(ASan)\cite{Asan} and flags all of them as gadgets.

\subsubsection{SpecTaint}
SpecTaint\cite{SpecTaint} is currently the only dynamic binary analysis tool for detecting Spectre gadgets and hence our direct competitor.
Its program instrumentation mechanisms are built on top of DECAF\cite{DECAF}, an extension of the full-system emulator QEMU\cite{QEMU}.
SpecTaint introduces dynamic information flow tracking~(DIFT)\cite{DIFT} to mark user inputs and trace the information flow to detect gadgets.

\subsubsection{Kasper}
Kasper\cite{Kasper} takes a principled approach and currently achieves the highest coverage in terms of gadget detection ability.
It is specifically built for testing the Linux kernel and adds instrumentation during compilation (and thus requires source code).
To detect attacker controllability and identify secrets, Kasper tracks the data flow by employing DIFT and implements kernel support for the DataFlowSanitizer\cite{DFSan} of LLVM.\
It also utilizes ASan to identify out-of-bounds accesses.
It considers both user-controlled input and speculatively uncontrolled out-of-bounds access to potentially load a secret.
In addition to the cache side channel, it also includes microarchitectural data sampling~(MDS)\cite{RIDL,Fallout} and port contention channels.
\textit{We adopt the Kasper policy to \toolname{} in this work for its prominent detection coverage and accuracy, further described in \cref{ssec:nahco3-gadget}.} \looseness=-1

%% file: figs/lst-specfuzz.tex
\begin{figure}[t]
\begin{lstlisting}[style=my-lst-style-linenum, caption={Simulation of the behavior of a program during branch misprediction (transient execution). Colored lines show added instrumentation.}, label={lst:specfuzz}]
void victim(int index) { // index == 20
@\lstcolourline{blue!15}@  if (!in_simulation) {
@\lstcolourline{blue!15}@    make_checkpoint();
@\lstcolourline{blue!15}@    if (index < 10) goto B; 
@\lstcolourline{blue!15}@    else goto A;@\tikzmark{specfuzz_goto_a}@
@\lstcolourline{blue!15}@  }
  @\tikzmark{specfuzz_if}@if (index < 10) {
A:  secret = foo[index];@\tikzmark{specfuzz_a}@
    baz = bar[secret];
@\lstcolourline{blue!15}@    @\tikzmark{specfuzz_restore}@if (in_simulation) rollback();
  }
B:return;
}
\end{lstlisting}
\begin{tikzpicture}[remember picture, overlay]
\draw[thick, arrows = {-Stealth[inset=0pt, angle=45:5pt]}, black] 
    ([shift={(5pt, 1.5pt)}] pic cs:specfuzz_goto_a)
    -| ([shift={(20pt, 1.5pt)}] pic cs:specfuzz_a)
    -- ([shift={(5pt, 1.5pt)}] pic cs:specfuzz_a);

\draw[thick, arrows = {-Stealth[inset=0pt, angle=45:5pt]}, black] 
    ([shift={(-5pt, 1.5pt)}] pic cs:specfuzz_restore)
    -- ([shift={(-23pt, 1.5pt)}] pic cs:specfuzz_restore)
    |- ([shift={(-2pt, 2.5pt)}] pic cs:specfuzz_if);
\end{tikzpicture}
\end{figure}

%% file: threat-model.tex
\subsection{Threat Model}

We assume the source code of the target program is missing.
Except for this assumption, we share a similar threat model with various other Spectre gadget detection studies\cite{SpecFuzz,SpecTaint,Kasper}.
We assume that the application is a potential victim of speculative information leaks.
To simplify the problem setting, we aim to detect Spectre-V1 gadgets leaking through the cache, MDS, and port contention side channels.
Although the proposed system can be extended to support other variants, most of them already have production-grade mitigations\cite{IntelSpectreMitigations}; in contrast, Spectre-V1 still cannot be easily mitigated with relatively low overhead.
We do not attempt to detect non-Spectre type vulnerabilities (e.g.,~memory safety violations) and assume they do not exist in the target application.
Following a previous study\cite{SpecFuzz}, we also do not attempt to detect Spectre gadgets guarded by more than six branches (including the first branch misprediction) as such gadgets are unlikely to be exploitable in the real world.

%% file: motivation.tex
\section{Motivation}\label{sec:motivation}
\input{figs/background-runtime-overhead}
\input{figs/switch-compiled-example}
\input{figs/tool-flowchart}

\subsection{Inefficiencies of full-system emulation}
As discussed earlier, the only available dynamic binary analysis tool for detecting Spectre gadgets is SpecTaint\cite{SpecTaint}.
Speaking of practicality, the problem with SpecTaint lies in its QEMU-based implementation; it inevitably limits its run-time efficiency, ultimately hindering gadget detection ability.
We demonstrate this by comparing the run-time performance of SpecTaint vs.\ SpecFuzz (a compiler-based approach).
We instrument two test programs, namely \texttt{libyaml} and \texttt{jsmn} and evaluate their execution times with crafted large inputs.
For a fair comparison, we disable their nested speculation simulation (described further in \cref{ssec:nahco3-ckpt-support}) and heuristic simulation skipping mechanisms.
The results are shown in \cref{fig:background-runtime-overhead}.
While SpecFuzz, an LLVM-based analysis tool, inevitably shows a large performance overhead (${\approx}600\times$ and ${\approx}1,800\times$) mainly due to speculation simulation, SpecTaint is another order of magnitude slower: compared to SpecFuzz, it is 28.5$\times$ and 11.1$\times$ slower on \texttt{libyaml} and \texttt{jsmn}, respectively.\looseness=-1

Due to its large performance overhead, SpecTaint trades off detection ability and only simulates the misprediction of each branch up to five times.
This leaves many potential false negatives on the table (details in \cref{ssec:exp-fuzz-v2}).
In addition, being based on a system emulator, SpecTaint lacks knowledge of critical program data structures such as the heap and stack.
It also cannot differentiate out-of-bounds and legal access.
Therefore, it has to be overly restrictive and assume all user-controlled memory accesses load secrets, resulting in many false positives, as shown in a previous study\cite{Kasper}.\looseness=-1

\subsection{Pitfalls of compiler-based approaches}\label{ssec:motivations-compiler}
Since both SpecFuzz and Kasper leverage LLVM compiler passes for instrumenting programs, they do not align with our assumption.
However, for the sake of demonstrating \toolname{}'s effectiveness, we perform discussions and comparisons against SpecFuzz in terms of performance and gadget detection ability throughout this paper (Kasper cannot be compared since it is tailored for analyzing the Linux kernel).

Generally speaking, using source code and information from the compiler for instrumentation can be more effective than static binary rewriting.
However, our observations highlight overlooked aspects by previous works, bringing attention that compiler-based methods have their own limitations and require careful application.

First, we illustrate one example by compiling a \texttt{switch} statement with GCC and Clang.
The respective assembly instructions are shown in \cref{fig:switch-compiled-example}.
The statement may be compiled into a combination of branches or a jump table depending on the compiler; only the former may result in Spectre-V1 gadgets.
If the analysis is done with Clang (as in SpecFuzz or Kasper), the generated program will not contain any gadgets.
However, if the final deployed programs are compiled with GCC instead, some gadgets might go undetected.
This means that compiler instrumentation-based approaches can be prone to false negatives and false positives due to the mismatch of compilers (and/or optimization levels) used for analysis vs.\ deployment.

We make another point that compiler-based methods can be inaccurate due to instrumentations commonly taking place before optimizations.
As a concrete example, SpecFuzz relies on ASan (to detect out-of-bounds accesses), whose instrumentations happen at the compiler \textit{frontend}.
It also implements most of its instrumentation passes in the compiler \textit{backend}.
Since its speculation simulation instrumentations happen in the \textit{backend}, it cannot distinguish the original program code vs.\ ASan instrumented code.
This hurts run-time performance (and hence gadget detection ability) since instrumentations are unintentionally added to ASan code; to make matters worse, ASan code is counted as program instructions, rendering the length of transient execution simulation (until the reorder buffer fills) inaccurate.
Porting everything into the compiler backend will potentially solve the problem, but requires huge engineering effort.\looseness=-1

%% file: figs/background-runtime-overhead.tex
\begin{figure}[t]
\scalebox{0.90}{
\begin{tikzpicture}%

\begin{axis}[
scaled ticks=false,
name=overheadfig,
nodes near coords,
nodes near coords align={horizontal},
nodes near coords style={anchor=west, font={\small}},
point meta=explicit symbolic,
axis y line*=left, x axis line style={draw=none, font={\small}},
xmajorgrids=true, xminorgrids=true, minor tick num=1,
major grid style={line width=.6pt,draw=gray!60},
xbar=0pt, bar width=.4cm, ytick=data, xmin=0, xmax=21500,
width=0.48\textwidth, height=4cm,
xlabel={Normalized Run Time},
x label style={yshift=1mm},
xticklabel={\pgfmathprintnumber{\tick}$\times$},
xtick style={yshift=1mm},
symbolic y coords={jsmn, libyaml},
yticklabel style={rotate=90},
ytick style={draw=none}, y axis line style={line width=1pt,draw=black},
enlarge y limits=0.5,
legend style={at={(1, 1)}, anchor=north west, font=\footnotesize},
legend image code/.code={\draw [#1] (0cm,-0.1cm) rectangle (0.2cm,0.15cm); },
reverse legend,
]
\addplot[fill=black!55]
coordinates {
    (1809,jsmn) [1809$\times$]
    (595,libyaml) [595$\times$]
};

\addplot[fill=black!85]
coordinates {
    (20196,jsmn) [20196$\times$]
    (16985,libyaml) [16985$\times$]
};

\legend{SpecFuzz, SpecTaint}
\end{axis}

\end{tikzpicture}
}
\caption{Comparison of the execution time of two programs instrumented with SpecTaint and SpecFuzz, with large crafted inputs, normalized to the execution time of the native versions. Shorter is better. Averaged over 10 runs.}\label{fig:background-runtime-overhead}
\end{figure}
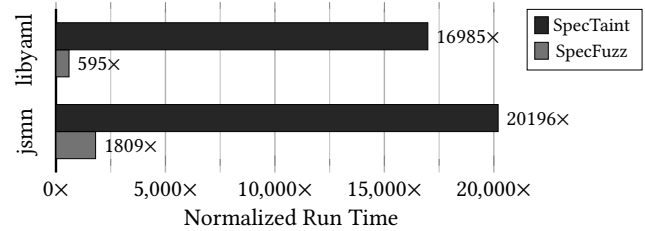

%% file: figs/switch-compiled-example.tex
\begin{figure*}[t]

\begin{tcbraster}[raster columns=10, enhanced, sharp corners, listing only, colback=white, boxrule=1pt, raster equal height, left=5pt, right=5pt, top=0pt, bottom=0pt]
\begin{tcblisting}{title=Program Source, listing options={style=my-lst-style-fullwidth}, raster multicolumn=3}
switch (value) {
  case 0: /*...*/ break;
  case 1: /*...*/ break;
  case 2: /*...*/ break;
  case 3: /*...*/ break;
}
\end{tcblisting}
\begin{tcblisting}{title=Compiled (GCC 13), listing options={style=my-lst-style-fullwidth, language={[x64]Assembler}}, raster multicolumn=3}
@\tikzmark{switch_gcc_northwest}@cmp edi, 2
@\lstcolourline{red!15}@je .L2
cmp edi, 3
@\lstcolourline{red!15}@je .L3
cmp edi, 1
@\lstcolourline{red!15}\tikzmark{switch_gcc_southwest}@je .L1
\end{tcblisting}
\begin{tcblisting}{title=Compiled (Clang 17), listing options={style=my-lst-style-fullwidth, language={[x64]Assembler}}, raster multicolumn=4}
@\tikzmark{switch_clang_northwest}@mov eax, edi
lea rcx, [rip + .LJT]
movsxd rax, dword ptr [rcx+4*rax]
add rax, rcx
@\lstcolourline{green!15}@jmp rax
@\tikzmark{switch_clang_southwest}@
\end{tcblisting}
\end{tcbraster}

\definecolor{alizarin}{rgb}{0.82, 0.1, 0.26}
\definecolor{dartmouthgreen}{rgb}{0.05, 0.5, 0.06}

\begin{tikzpicture}[remember picture, overlay]
\draw[alizarin, line width=.1cm, fill=white] ([shift={(4.2cm, .3cm)}] pic cs:switch_gcc_southwest) circle (.5cm) node (specv1_vuln) {\includegraphics[width=.7cm]{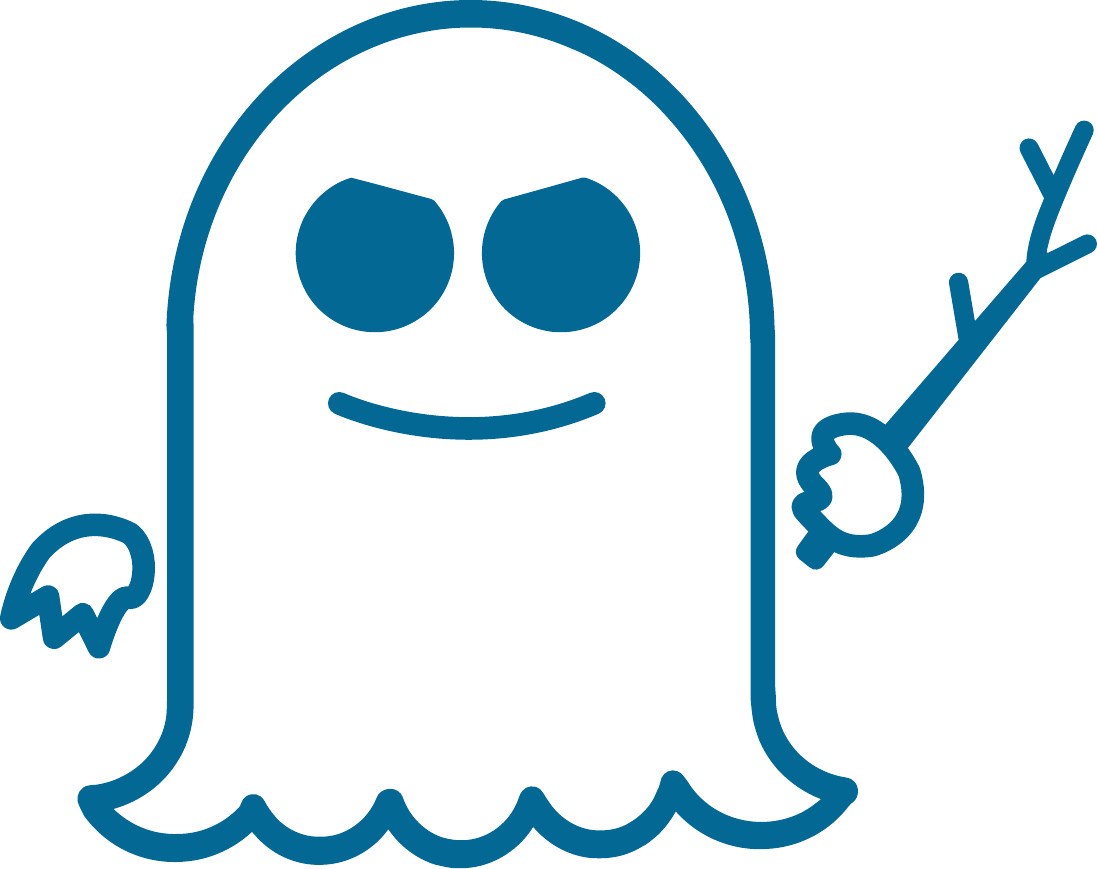}};

\node[draw,align=center,fill=white] at ([shift={(3.8cm, 0cm)}] pic cs:switch_gcc_northwest) {Spectre-V1\\\textbf{Vulnerable}};

\draw[dartmouthgreen, line width=.1cm, fill=white] ([shift={(6cm, .3cm)}] pic cs:switch_clang_southwest) circle (.5cm) node [yshift=-.02cm] (specv1_safe) {\includegraphics[width=.7cm]{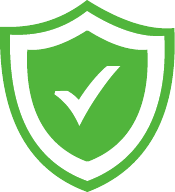}};

\node[draw,align=center,fill=white] at ([shift={(5.7cm, 0cm)}] pic cs:switch_clang_northwest) {Spectre-V1\\\textbf{Safe}};
\end{tikzpicture}

\caption{Assembly instructions of a switch statement compiled with GCC and Clang, respectively, with \texttt{-O2} optimization. The different structures of the generated code affect the existence of Spectre-V1 gadgets.}\label{fig:switch-compiled-example}
\end{figure*}

%% file: figs/tool-flowchart.tex
\begin{figure*}[th]
\centering

\tikzstyle{flowchartnode} = [rectangle, rounded corners, minimum width=1.5cm, minimum height=1cm, text centered, draw=black, fill=gray!15, align=center, font={\fontsize{8pt}{9}\selectfont}]
\tikzstyle{flowchartinnernode} = [rectangle, rounded corners, text centered, draw=black, fill=white, align=center, font={\fontsize{8pt}{9}\selectfont}]
\tikzstyle{flowchartnote} = [font={\fontsize{7pt}{7}\selectfont}, anchor=south, align=center]

\scalebox{0.95}{
\begin{tikzpicture}[node distance=3.3cm]

\node (binary) [flowchartnode] {Binary\\Program};
\node (asm) [right of=binary, flowchartnode] {Assembly};
\node (normal_copy) [right of=asm, yshift=.6cm, flowchartnode, minimum width=1.8cm] {\normalfnname{}};
\node (spec_copy) [right of=asm, yshift=-.6cm, flowchartnode, minimum width=1.8cm] {\specfnname{}};
\node (normal_copy_inst) [right of=normal_copy, flowchartnode] {\textsc{Instrumented}\\\normalfnname{}};
\node (spec_copy_inst) [right of=spec_copy, flowchartnode] {\textsc{Instrumented}\\\specfnname{}};
\node (inst_binary) [right of=normal_copy_inst, yshift=-.6cm, flowchartnode, minimum height=2.2cm] {\textsc{Instrumented}\\Program\\\\\\\\};
\node (inst_binary_real_copy) [flowchartinnernode, minimum width=1.8cm] at ([shift={(0, -.2cm)}] inst_binary.center) {Real Copy};
\node (inst_binary_shadow_copy) [flowchartinnernode, minimum width=1.8cm] at ([shift={(0, -.8cm)}] inst_binary.center) {Shadow Copy};
\node (gadgets) [right of=inst_binary, flowchartnode, fill=white] {\includegraphics[width=.8cm]{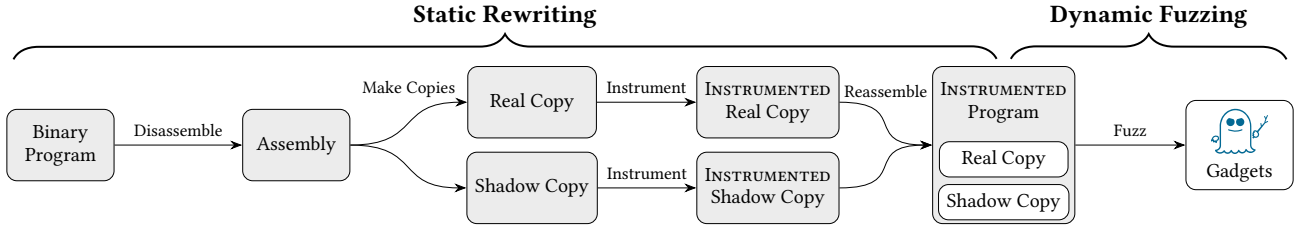}\\Gadgets};

\draw [-Stealth] (binary) -- node[flowchartnote] {Disassemble} (asm);
\draw [-Stealth] (asm) .. controls ([shift={(1cm, 0)}] asm.east) and ([shift={(-1cm, 0)}] normal_copy.west) .. 
    node[flowchartnote, yshift=.25cm] {Make Copies} (normal_copy);
\draw [-Stealth] (asm) .. controls ([shift={(1cm, 0)}] asm.east) and ([shift={(-1cm, 0)}] spec_copy.west) .. (spec_copy);
\draw [-Stealth] (normal_copy) -- node[flowchartnote] {Instrument} (normal_copy_inst);
\draw [-Stealth] (spec_copy) -- node[flowchartnote] {Instrument} (spec_copy_inst);
\draw [-Stealth] (normal_copy_inst) .. controls ([shift={(1cm, 0)}] normal_copy_inst.east) and ([shift={(-1cm, 0)}] inst_binary.west) ..
    node[flowchartnote, yshift=.25cm] {Reassemble} (inst_binary);
\draw [-Stealth] (spec_copy_inst) .. controls ([shift={(1cm, 0)}] spec_copy_inst.east) and ([shift={(-1cm, 0)}] inst_binary.west) .. (inst_binary);
\draw [-Stealth] (inst_binary) -- node[flowchartnote] {Fuzz} (gadgets);

\draw [thick, decorate, decoration = {brace, amplitude=10pt}] ([shift={(.1cm, 1.2cm)}] binary.west) -- 
    node[anchor=south, yshift=.3cm] {\textbf{Static Rewriting}} ([shift={(-.1cm, .1cm)}] inst_binary.north);
\draw [thick, decorate, decoration = {brace, amplitude=10pt}] ([shift={(.1cm, .1cm)}] inst_binary.north) -- 
    node[anchor=south, yshift=.3cm] {\textbf{Dynamic Fuzzing}} ([shift={(-.1cm, 1.2cm)}] gadgets.east);

\end{tikzpicture}
}
\caption{Workflow of detecting Spectre gadgets in binaries with \toolname{}.}\label{fig:tool-flowchart}
\end{figure*}

%% file: overview.tex
\section{\toolname{}: Overview}

To the best of our knowledge, \toolname{} is the first static-binary-rewriting-based Spectre gadget detector.
It accepts a user-space COTS binary program as an input and works without the presence of source code.

\cref{fig:tool-flowchart} presents the workflow of \toolname{}, which can be divided into two stages.
In the first static rewriting stage, \toolname{} fits the binaries with static instrumentation that helps detect Spectre gadgets.
The second stage relies on existing techniques to perform dynamic fuzzing where the binaries are fuzzed with the instrumentation that dynamically detects gadgets at run time.

In the following sections, we first present \twocopyname{}, our fundamental design concept that splits the program into Real Copy and Shadow Copy for enhancing performance in \cref{sec:twocopy}. %
Next, we introduce the architecture and implementation of \toolname{} in \cref{sec:architecture}.

%% file: two-copy-approach.tex
\section{\twocopyname}\label{sec:twocopy}

\subsection{Problem analysis}\label{ssec:twocopy-motivation}

\begin{lstlisting}[style=my-lst-style-linenum, caption=Mixing normal and speculation simulation code adds the program with unnecessary instrumentation and guard conditionals., label={lst:inst-without-two-copy}, float=tp]
@\lstcolourline{blue!15}@if (!in_simulation) start_simulation();
if (index < SIZE) {
@\lstcolourline{green!15}@  if (in_simulation) asan_check();
@\lstcolourline{green!15}@  if (in_simulation) write_memlog();
  secret = foo[index];
@\lstcolourline{green!15}@  if (in_simulation) asan_check();
@\lstcolourline{green!15}@  if (in_simulation) write_memlog();
  baz = bar[secret];
@\lstcolourline{green!15}@  if (in_simulation) end_simulation();
}
\end{lstlisting}

\cref{lst:inst-without-two-copy} shows the common architecture of previous studies on simulating transient execution.
The program:
\begin{enumerate}
  \item checkpoints itself and switches from the normal execution mode to the speculation simulation mode before every conditional branch (line 1);
  \item performs ASan check upon \textit{every memory access} and logs \textit{every memory update} in order to restore the memory state upon rollback (lines 3--4 and 6--7);
  \item executes some instructions to simulate transient execution (lines 3--8); rolls back to the checkpoint and restores the state (line 9); and continues normal execution.
\end{enumerate}
The problem is that different types of instrumentations for the two modes coexist in the program to detect Specter gadgets: instrumentation for creating checkpoints and flipping conditional branches in the normal execution mode; ASan checks, memory logging and rollback points in the speculation simulation mode.
All the instrumentation codes are added to the same binary, guarded by conditional branches checking if the current mode matches what the instrumentation is intended for.\looseness=-1

Obviously, this form of instrumentation is not ideal because the guard condition \texttt{if (in\_simulation)} must be checked for \textit{each} instrumentation at run time.
Roughly speaking around 25\% of dynamically executed instructions are loads (require ASan check) and 10\% are stores (require ASan check and memory logging) in integer programs\cite{HennessyPatterson}.
In addition, there are other forms of instrumentation added to the program (explained later).
The above factors combined significantly slow down the program execution (and worsen fuzzing efficiency).

The key insight of \twocopyname{} is that, by completely separating the binary code executed during normal execution and speculation simulation, we can equip them with only the needed instrumentation types.
This eliminates the need for conditional guards around each instrumentation, making the instrumented binary efficient and performant. \looseness=-1

\subsection{Illustrative demonstration}
\input{figs/two-copy-demo}

We leverage reassembleable disassembly techniques\cite{Ddisasm} to statically rewrite binaries.
Therefore, the source code examples shown in this section are not required for \toolname{} and are solely used for illustration purposes.

An overview of \twocopyname{} is shown in \cref{fig:two-copy-demo}.
During instrumentation time, we make a complete, byte-to-byte copy of each function, and append its name with a fixed suffix (we use \texttt{\$spec} for this purpose).
We refer to the copied function as the \specfnname{} and the original function as the \normalfnname{}.
Then, we update all the control flow transitions that are known at compile time, namely direct branches and function calls, to instead refer to their counterparts in the \specfnname{} (indirect branch handling is discussed in \cref{ssec:twocopy-ind-call}).
These measures ensure that the control flow never escapes unexpectedly into code that does not correspond to the current execution mode.
Then, we populate the two copies with their corresponding sets of instrumentation, which can remove all the unnecessary guard conditionals.

When the program runs in normal execution mode and encounters a conditional branch, the program starts to simulate branch misprediction (transient execution).
It enters speculation simulation mode by calling \texttt{start\_simulation}.
The function takes a checkpoint of the current program state and transfers the control flow to the corresponding destination of the reversed branch outcome in the \specfnname{}.

We create trampolines for each conditional branch during instrumentation to facilitate this transition (shown in the bottom middle box labeled ``Trampoline'' in \cref{fig:two-copy-demo}).
Each trampoline consists of two jump instructions.
The first jump follows the same condition as the original branch instruction; but instead of going to the original jump target (label A in the ``Shadow Copy''), we set the jump target to the opposite destination (label B; the destination when the condition is false) in the \specfnname{}.
Then, the second jump is unconditional and brings the control flow to the original jump target in the \specfnname{}.
The program therefore follows the trampoline and enters the \specfnname{} on the deliberately-made incorrect execution path, effectively starting the simulation for the transient execution following a branch misprediction. \looseness=-1

The program continues to execute until it encounters a restore point \texttt{end\_simulation}.
If the program satisfies one of the termination conditions, e.g., the number of instructions the program has executed in the \specfnname{} is greater than a preset threshold (to simulate the processor no longer speculates because the reorder buffer is full; we use 250 as in prior studies), a rollback is initiated.
It completely resets the program to the state before the simulation was started, and continues normal execution.

In the example depicted in \cref{fig:two-copy-demo}, the function \texttt{fib} calculates the $i$-th Fibonacci number if $i$ is within the array bounds, or returns -1 as an error otherwise.
Assume that we provide an \texttt{i} such that \mbox{\texttt{i >= SIZE}}, so on the correct execution path, the function returns -1.
However, after the call to \texttt{start\_simulation}, the program enters the trampoline, and instead of going to target B, it enters target A in \specfnname{} that calculates the next Fibonacci number, and attempts to access \texttt{f[i-1]} and \texttt{f[i-2]} which are both out of bounds.
These illegal accesses are caught by the ASan instrumentation (and are reported to the fuzzing runtime).

\subsection{Handling returns and indirect calls/jumps}\label{ssec:twocopy-ind-call}

\input{figs/control-flow-escape}
\input{figs/lst-ind-br-patch}

It is important that the control flow never goes into the wrong function copy.
This is not an issue for the \normalfnname{} code, as all the execution side effects from the \specfnname{} are discarded during rollback, so code pointers from the speculation mode never leak backward into the normal execution mode.
On the other hand, a few cases exist in which a program executing in speculation simulation mode may accidentally jumps into \normalfnname{} code.
When a \normalfnname{} code pointer is handled in the \specfnname{} code, such as through function returns, indirect calls and jumps etc, the control flow may no longer be confined in the \specfnname{} and escapes.
This is highly undesirable, because the speculation mode will never terminate as no restore points are inserted into the \normalfnname{}, and the program will continue to execute in an incorrect state.\looseness=-1

\cref{fig:control-flow-escape} shows two examples:
\begin{enumerate}
  \item In \cref{fig:control-flow-escape-return}, function \texttt{run\_fib} calls \texttt{fib}, and the program enters speculation simulation and jumps to \texttt{fib\${\allowbreak}spec}.
  When \texttt{fib\$spec} returns (still in speculation simulation), it does not stay in the \specfnname{} but goes straight back to \texttt{run\_fib} in the \normalfnname{}.
  \item In \cref{fig:control-flow-escape-pointer}, a function pointer of \normalfnname{} is taken as a global variable and passed into \specfnname{}, which is later called to escape from speculation simulation. \looseness=-1
\end{enumerate}
Since indirect branches cannot be resolved at compile time, we must actively detect and redirect such unexpected control flow transitions at run time.
We apply special transformations to the functions as shown in \cref{lst:ind-br-patch}.

First, we detect if a basic block in the \normalfnname{} may be a target of an indirect control flow transfer at instrumentation time. %
Indirect control flow transfers include returns from a function call (such as \cref{fig:control-flow-escape-return}), indirect calls and indirect jumps.

For all such targets (basic blocks), we insert a special marker \texttt{nop} instruction at the start of the basic block in the \normalfnname{} (line~12), one that compilers do not normally generate. %
Then, we add checks for if the current run-time mode is in speculation simulation to detect control-flow escapes, in which case we redirect the control flow back by inserting an unconditional jump to the counterpart of the current basic block in the \specfnname{} (lines~13--14).

For all control flow instructions that involve a code pointer in the \specfnname{}, we check if the code pointer belongs to either the \specfnname{} or the \normalfnname{} (line~2).
In the latter case, we also check if the jump target is a transformed target that can detect control flow escapes by checking for the existence of the special marker \texttt{nop} instruction (lines~4--5).
If the check passes, the indirect branch proceeds; otherwise, a forced rollback is launched that terminates the simulation to not corrupt the control flow (line~8).

Admittedly, this method adds run-time overheads to maintain the integrity of the control flow, which seems to (somewhat) diminish the purpose of \twocopyname{}.
However, we argue that such overhead is acceptable.
This is because indirect control transfers (function calls, returns, and indirect calls and jumps) are relatively costly, and the instrumentation we add brings little additional overhead.
Compilers also tend to inline frequent function calls when the programs are compiled with high optimization levels, so in a well-optimized program, the number of these added checks is likely non-significant. 

\vspace{\baselineskip}

%% file: figs/two-copy-demo.tex
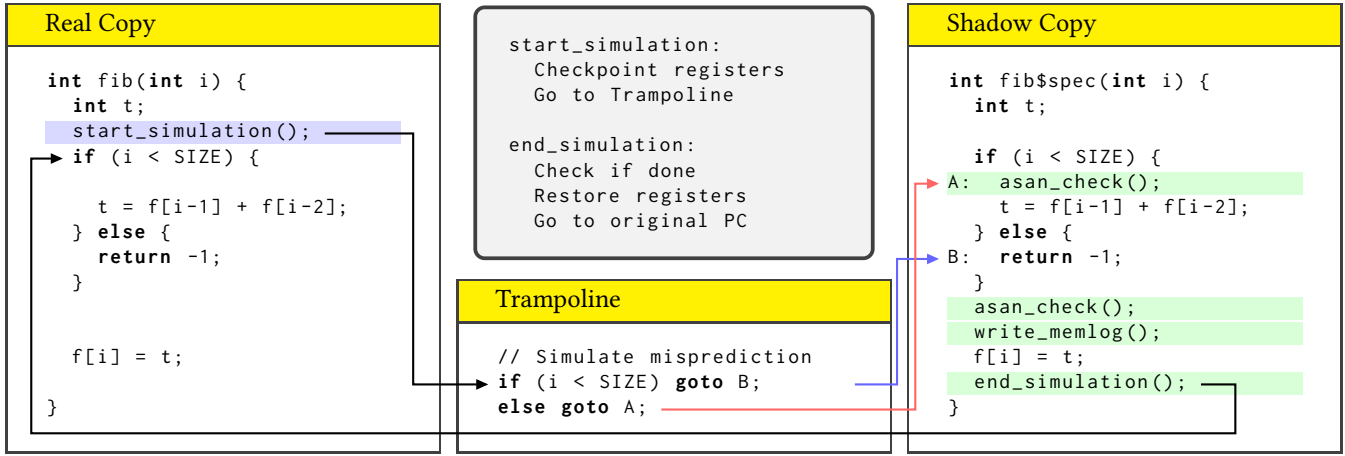
\begin{figure*}[t]
\begin{tcbraster}[raster columns=3, raster valign=bottom, enhanced, sharp corners, listing only, colbacktitle={yellow}, coltitle=black, colback=white, boxrule=1pt, top=2pt]
\begin{tcblisting}{title=\normalfnname, listing options={style=my-lst-style-fullwidth}}
int fib(int i) {
  int t;
@\lstcolourline{blue!15}@  start_simulation(); @\tikzmark{twocopy_ckpt}@
  @\tikzmark{twocopy_ckpt_next}@if (i < SIZE) {
  
    t = f[i-1] + f[i-2];
  } else {
    return -1;
  }

  f[i] = t;
  
}
\end{tcblisting}
\begin{tcblisting}{title=Trampoline, listing options={style=my-lst-style-fullwidth}}
// Simulate misprediction
@\tikzmark{twocopy_trampoline}@if (i < SIZE) goto B; @\tikzmark{twocopy_trampoline_b}@
else goto A; @\tikzmark{twocopy_trampoline_a}@
\end{tcblisting}
\begin{tcblisting}{title=\specfnname, listing options={style=my-lst-style-fullwidth}}
int fib@\$@spec(int i) {
  int t;

  if (i < SIZE) {
@\lstcolourline{green!15}\tikzmark{twocopy_a}@A:  asan_check();
    t = f[i-1] + f[i-2];
  } else {
@\tikzmark{twocopy_b}@B:  return -1;
  }
@\lstcolourline{green!15}@  asan_check();
@\lstcolourline{green!15}@  write_memlog();
  f[i] = t;
@\lstcolourline{green!15}@  end_simulation(); @\tikzmark{twocopy_restore}@
}
\end{tcblisting}
\end{tcbraster}

\begin{center}
\begin{tikzpicture}[remember picture,overlay]
\node[] at (0, 127pt) {
\begin{tcolorbox}[width=0.3\linewidth, boxsep=0pt]
\begin{lstlisting}[style=my-lst-style-fullwidth,language=none]
start_simulation:
  Checkpoint registers
  Go to Trampoline

end_simulation:
  Check if done
  Restore registers
  Go to original PC
\end{lstlisting}
\end{tcolorbox}
};

\draw[thick, arrows={-Stealth[inset=0pt, angle=45:5pt]}, black] 
    ([shift={(0, 1.5pt)}] pic cs:twocopy_ckpt)
    -- ([shift={(5pt, 1.5pt)}] pic cs:twocopy_ckpt)
    -|- ([shift={(-3pt, 1.5pt)}] pic cs:twocopy_trampoline);
\draw[thick, arrows = {-Stealth[inset=0pt, angle=45:5pt]}, blue!60] 
    ([shift={(30pt, 1.5pt)}] pic cs:twocopy_trampoline_b)
    -|- ([shift={(-3pt, 1.5pt)}] pic cs:twocopy_b);
\draw[thick, arrows = {-Stealth[inset=0pt, angle=45:5pt]}, red!60] 
    ([shift={(0, 1.5pt)}] pic cs:twocopy_trampoline_a)
    -- ([shift={(87pt, 1.5pt)}] pic cs:twocopy_trampoline_a)
    -|- ([shift={(-3pt, 1.5pt)}] pic cs:twocopy_a);
\draw[thick, arrows = {-Stealth[inset=0pt, angle=45:5pt]}, black] 
    ([shift={(0, 1.5pt)}] pic cs:twocopy_restore)
    -- ([shift={(13pt, 1.5pt)}] pic cs:twocopy_restore)
    -- ([shift={(13pt, -17pt)}] pic cs:twocopy_restore)
    -| ([shift={(-15pt, 1.5pt)}] pic cs:twocopy_ckpt_next)
    -- ([shift={(-3pt, 1.5pt)}] pic cs:twocopy_ckpt_next);
\end{tikzpicture}
\end{center}

\caption{Flow of program execution and misspeculation simulation with \twocopyname{}. } %
\label{fig:two-copy-demo}

\end{figure*}

%% file: figs/control-flow-escape.tex
\begin{figure}[t]
\centering
\begin{subfigure}[b]{0.38\columnwidth}
\centering
\begin{tikzpicture}[
    level/.style={draw, minimum width=3cm, minimum height=6.7mm},
    stack/.style={matrix of nodes, nodes={level}, row sep=-\pgflinewidth}
]
\matrix[stack, label=Call Stack] at (0,0) (call_stack) {
    fib\$spec+0x114 \\
    run\_fib+0x514 \\
    main+0x810 \\
};
\end{tikzpicture}
\caption{Function Return}
\label{fig:control-flow-escape-return}
\end{subfigure}\hspace{0.06\columnwidth}
\begin{subfigure}[b]{0.55\columnwidth}
\centering
\begin{lstlisting}[language=c, style=my-lst-style, linewidth=\columnwidth]
fn *bar = &nonspec_fn;
void foo$spec(fn *baz) {
  baz();
}

int main$spec() {
  foo$spec(bar);
}
\end{lstlisting}
\caption{Explicit Pointer Usage}
\label{fig:control-flow-escape-pointer}
\end{subfigure}
\caption{Examples of control flow escaping into \normalfnname{}.}
\label{fig:control-flow-escape}
\end{figure}
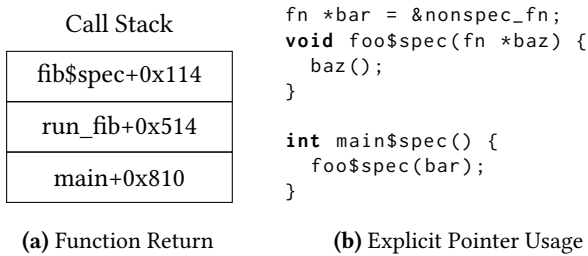

%% file: figs/lst-ind-br-patch.tex
\begin{figure}[t]
\begin{lstlisting}[style=my-lst-style-linenum, caption={Indirect branch and branch target transformations to prevent control flow escapes.}, label={lst:ind-br-patch}]
void foo$spec() { // ...
@\lstcolourline{green!15}@  if (in_shadow_copy(ret_addr))
@\lstcolourline{green!15}@    return;
@\lstcolourline{green!15}@  else if (in_real_copy(ret_addr) &&
@\lstcolourline{green!15}@    has_special_nop_marker(ret_addr))
@\lstcolourline{green!15}@    return;@\tikzmark{indbr-patch_ret}@
@\lstcolourline{green!15}@  else
@\lstcolourline{green!15}@    end_simulation();
}
void bar() { // ...
  foo();
@\lstcolourline{blue!15}@  special_nop_marker();@\tikzmark{indbr-patch_rettarget}@
@\lstcolourline{blue!15}@  if (in_simulation) 
@\lstcolourline{blue!15}@    goto AFTER_FOO_IN_SPEC;@\tikzmark{indbr-patch_goto}@
}
void bar$spec() { // ...
  foo$spec();
@\lstcolourline{green!15}@AFTER_FOO_IN_SPEC: // ...@\tikzmark{indbr-patch_gototarget}@
}
\end{lstlisting}

\begin{tikzpicture}[remember picture, overlay]
\draw[thick, arrows = {-Stealth[inset=0pt, angle=45:5pt]}, black] 
    ([shift={(5pt, 1.5pt)}] pic cs:indbr-patch_ret)
    -| ([shift={(20pt, 1.5pt)}] pic cs:indbr-patch_rettarget)
    -- ([shift={(5pt, 1.5pt)}] pic cs:indbr-patch_rettarget);

\draw[thick, arrows = {-Stealth[inset=0pt, angle=45:5pt]}, black] 
    ([shift={(5pt, 1.5pt)}] pic cs:indbr-patch_goto)
    -- ([shift={(20pt, 1.5pt)}] pic cs:indbr-patch_goto)
    |- ([shift={(5pt, 1.5pt)}] pic cs:indbr-patch_gototarget);
\end{tikzpicture}

\end{figure}

%% file: tool-arch.tex
\section{\toolname{}: Architecture and Implementation}\label{sec:architecture}

We employ \twocopyname{} to implement \toolname{}, the first Spectre gadget detector based on static binary rewriting.
\toolname{} is built on the state-of-the-art binary disassembly and rewriting platform Datalog Disassembly\cite{Ddisasm} and GTIRB\cite{GTIRB}.
It is implemented with $\sim$2,000 lines of Python and $\sim$1,000 lines of C and x86 assembly.
While the current implementation supports unobfuscated x86-64 Linux ELF binaries, the design of \toolname{} is not specific to any instruction set or operating system and can be easily extended.\looseness=-1

\subsection{Checkpoint and restore}\label{ssec:nahco3-ckpt-support}

We implement the following functions (which are akin to what is proposed in SpecFuzz\cite{SpecFuzz}) to facilitate program state checkpoints and rollback:

\textbf{Checkpoint.}
Before entering a speculation simulation, we pack and store general-purpose registers, flags, and the program counter as a checkpoint.
We also store the SSE registers by default, and only preserve the full AVX registers when a corresponding option is enabled for performance reasons.
After the checkpoint is created, the program enters the \specfnname{} through the trampoline.

\textbf{Memory Log.}
We do not directly checkpoint the memory since it is extremely inefficient to copy all the live memory.
Instead, we instrument all the memory write instructions in the \specfnname{} to first log the write address and the original contents into a memory log beforehand. \looseness=-1

\textbf{Rollback.}
During rollback, the logged memory contents are written back in reverse.
The registers are restored, and the program is returned to the previous program counter in the checkpoint.
With all side effects cleaned up, the program returns to the \normalfnname{} and continues normal execution. \looseness=-1

\textbf{Conditional Restore Points.}
The number of instructions that can be speculatively executed depends on the size of the reorder buffer.
To simulate this, we count the instructions executed in \specfnname{}.
The counting instrumentation is placed near the end of each basic block and between every 50 instructions; the actual position may vary slightly as we attempt to pick an insertion point with most free registers.
If the instruction counter exceeds the threshold (250 instructions), simulation ends and rollback begins.\looseness=-1

\textbf{Unconditional Restore Points.}
There exist cases where speculation simulation cannot continue, and a rollback must start.
This includes unidentified indirect branch targets as discussed in \cref{ssec:twocopy-ind-call} as continuing the execution will break control flow integrity.
When the program encounters an external library call in the \specfnname{}, the simulation must also be terminated, as external calls to uninstrumented libraries may cause side effects that cannot be recovered.
The simulation also concludes before serializing instructions such as \texttt{lfence} and \texttt{cpuid} (because it will terminate speculative execution).
An instrumentation pass inserts unconditional restore points at these locations.

\textbf{Exceptions.}
Inevitably, some of the execution paths in the \specfnname{} lead to exceptions.
We register a custom signal handler and conservatively launch a rollback when triggered.

\textbf{Nested Speculation and fuzzing heuristic.}
Spectre gadgets may be guarded by multiple branch mispredictions.
We enable nested simulations by inserting \texttt{start\_simulation} calls before conditional branches in the \specfnname{}.
Even within the 250-instruction constraint, the simulation can diverge into millions of different paths\cite{SpecFuzz} (as one out of five instructions is a branch on average\cite{HennessyPatterson}), and the search space grows exponentially.
Previous works introduce heuristics to manage the nested exploration: SpecFuzz keeps track of the number of encounters per branch and gradually increases the depth of simulation as its encounter, up to the sixth order;
SpecTaint\cite{SpecTaint} instead performs depth-first speculation for nested branches, however, enters speculation simulation for each branch only up to five times.
We adopt a mixture of their approaches to maximize gadget detection ability.
As it is highly unlikely that exploitable Spectre gadgets are guarded by more than six branches, branch depth is simulated up to the sixth order for the first five runs of a branch, and the simulation follows the SpecFuzz heuristics afterward.\looseness=-1

\subsection{Gadget detection and reporting}\label{ssec:nahco3-gadget}

\input{figs/kasper-policy.tex}

The architecture of \toolname{} is decoupled from the gadget detection policies so that we can implement different policies of our choice.
We chose to demonstrate \toolname{} with the Kasper policy shown in \cref{fig:kasper-policy}. %
It detects gadgets by utilizing the LLVM implementations of the AddressSanitizer (ASan)\cite{Asan} and DataFlowSanitizer (DFSan)\cite{DFSan}.
We demonstrate the binary-based implementations of ASan and DIFT.
Moreover, we discuss optimization opportunities of these building blocks specific to \twocopyname{}.

\subsubsection{Binary ASan}\label{sssec:nahco3-asan}

\textit{\underline{Memory Poisoning.}}
Poisoning heap memory is trivial as linking with the ASan library automatically hooks \texttt{malloc} and \texttt{free} calls.
However, inserting red zones for stack and global objects after compilation is difficult due to the absence of object type and boundary information.
Therefore, we protect stacks at a stack-frame granularity by poisoning the shadow bytes of the return addresses.
Unfortunately, protecting global objects with binary rewriting is impractical as discussed in previous work\cite{RetroWrite}, and we leave them unprotected.

\textit{\underline{ASan Checks.}}
We insert ASan checks before memory accesses in the \specfnname{}, with inline assembly snippets that read and test the shadow memory.
We always allowlist the accesses that are based off \texttt{rsp} or \texttt{rbp} and have a constant offset, in order to allow functions like \texttt{\_\_builtin\_return{\textunderscore\allowbreak}address} to still work.
We can omit adding ASan checks to the \normalfnname{} to improve performance, thanks to the design of \twocopyname{}.

\subsubsection{Binary DIFT}\label{sssec:nahco3-dift}

\begin{table}[t]
  \centering
  \caption{User-accessible memory regions with ASan\cite{Asan}.}
  {
  \rowcolors{1}{}{gray90}
  \resizebox{0.45\textwidth}{!}{
    {\renewcommand{\arraystretch}{1.2}
    \begin{tabular}{c>{\hspace{1pc}} c>{\hspace{1pc}}c} %
      \toprule
      Name & Start & End \\
      \midrule
      HighMem & \texttt{0x1000'7fff'8000} & \texttt{0x7fff'ffff'ffff} \\
      LowMem & \texttt{0x0} & \texttt{0x7fff'7fff} \\
      \bottomrule
    \end{tabular}}\label{tab:asan}
  }
}
\end{table}

\begin{table}[t]
  \centering
  \caption{User-accessible memory and tag shadow regions with ASan and the data flow tracker.}
  {
  \rowcolors{1}{}{gray90}
  \resizebox{0.45\textwidth}{!}{
    {\renewcommand{\arraystretch}{1.2}
    \begin{tabular}{c>{\hspace{1pc}} c>{\hspace{1pc}}c} %
      \toprule
      Name & Start & End \\
      \midrule
      HighMem & \texttt{0x6000'0000'0000} & \texttt{0x7fff'ffff'ffff} \\
      HighTag & \texttt{0x4000'0000'0000} & \texttt{0x5fff'ffff'ffff} \\
      LowTag & \texttt{0x2000'0000'0000} & \texttt{0x2000'7fff'7fff} \\
      LowMem & \texttt{0x0} & \texttt{0x7fff'7fff} \\
      \bottomrule
    \end{tabular}}\label{tab:bin-dift}

  }
}
\end{table}

\textit{\underline{Tag Shadow.}}
We utilize shadow memory to store the tags (metadata) for data flow tracking.
ASan devotes some of the address space to its shadow memory, leaving two user-accessible memory regions (as shown in \cref{tab:asan}).
We borrow some area in the user-addressable memory to store the tags.
We reserve part of the HighMem for the tag shadow as shown in \cref{tab:bin-dift}.
The tag shadow has a byte-to-byte mapping to the user-accessible memory, and the address can be translated by flipping the 45th bit.
Each byte represents the set of tags the corresponding data byte holds, while a bit represents one tag.
This design is similar to DFSan\cite{DFSan}.

\textit{\underline{Taint Sources.}}
Kasper policy (cf. \cref{fig:kasper-policy}) requires marking all user inputs as \textit{attacker-directly controlled (User)}.
We provide wrappers for common user input functions such as \texttt{fread} and \texttt{fgets} and tag their output accordingly.
We also tag \texttt{argc} and \texttt{argv}.
Kasper policy also marks the outcomes of speculative out-of-bounds accesses as \textit{attacker-indirectly controlled (Massage)} as they potentially construct wild data pointers; we use ASan to detect them and mark them as such.

\textit{\underline{Tag Propagation.}}
Benefiting from \twocopyname{}, we propose using different implementations of tag propagation for normal execution and speculation simulation, respectively.
In the \specfnname{}, we generate and insert assembly snippets that propagate the tag before each instruction and log the tag changes for later rollback.
However, in the \normalfnname{}, the program execution and the tag propagation do not always need to be synchronized since there are no taint sinks, leaving room for optimization.
As such, we asynchronously update the tags only once per basic block.
For this purpose, we generate a list of LLVM IR expressions that compute the tag changes for each block.
The list of IR expressions is optimized and compiled into an assembly snippet using LLVM bindings.
We insert the snippet into an optimal position in the basic block that yields minimum register spilling.
This significantly improves the performance of propagating tags in the \normalfnname{}.

\textit{\underline{Taint Sinks.}}
Kasper policy marks a load value as \textit{secret} if it comes from an \textit{attacker-directly controlled} out-of-bounds access or any access composed from an \textit{attacker-indirectly controlled} pointer (as such wild pointers often violate program invariants).
Before each memory access in the \specfnname{}, we check if the memory address has one of the \textit{attacker} tags.
If the tag is \textit{attacker-direct}, we perform ASan checks to detect out-of-bounds access and promote the loaded value's tag to \textit{secret} upon detection.
If the tag is \textit{attacker-indirect}, we always promote the tag to \textit{secret}.

When the \textit{secret} value is loaded, it is immediately vulnerable to being leaked via MDS side channels, so an MDS gadget is reported.
If the \textit{secret} value is used to compose a dereferenced pointer, this constitutes a secret transmitter via the cache side channel, and a cache gadget is recorded.
If the \textit{secret} influences the outcome of a branch (i.e., any of the operands of the last instruction that modifies FLAGS before a branch is a \textit{secret}), the secret changes the execution flow of the program and can be transmitted via port contention, so we report a port contention gadget.

\subsubsection{Gadget Reporting}

The found gadgets are handed over to the fuzzer via a custom signal.
Then, the fuzzer makes records of them for further processing.
Alternatively, the gadgets are also logged in a format compatible with the SpecFuzz workflow scripts, which enables reusing existing tools for further analysis and patching.

\subsection{Coverage tracking}\label{ssec:nahco3-cov}

Unlike other software testing scenarios, Spectre gadget detection can define two types of coverage: that of normal execution and speculation simulation.
In order to increase the gadget detection ability, we need to maximize speculation simulation coverage, which inevitably requires to increase normal execution coverage as well.
Although the interaction between the two is unclear, we believe it is valuable to track them separately and perform as such.

The design of \toolname{} is decoupled with the fuzzer, and \toolname{} interfaces the binaries with SanitizerCoverage\cite{SanCov} which defines standard coverage tracing schemes.
The instrumented binaries are then compatible to be tested with any fuzzer that supports such interface; we demonstrate with \texttt{honggfuzz} in the experiments.%

For the normal coverage, we trace the coverage at every conditional branch before entering speculation simulation.
However, there are challenges associated with tracing coverage of speculation simulation.
The tracking requires calling the coverage function, which clobbers a large number of registers and has a non-negligible overhead.
This is because every time we encounter a conditional branch instruction in normal execution mode, we need to execute \textit{many} basic blocks in speculation simulation where each one of them calls the coverage function (assuming one out of five instructions is a branch\cite{HennessyPatterson}, we need to trace 50 basic blocks for the transiently executed 250 instructions).

We propose an optimization specific for tracing the speculation simulation coverage to reduce the run-time overhead.
We create a statistic of the basic block count during instrumentation time and insert guards into the binary.
We leverage the fact that each round of speculation simulation is guaranteed to only execute a finite number of instructions and traverse finite basic blocks.
Thus, when visiting a basic block in the \specfnname{}, we only note its guard ID in a buffer.
Then, the actual coverage is only updated lazily before the rollback begins to eliminate the register preservation overhead.
\looseness=-1

%% file: figs/kasper-policy.tex
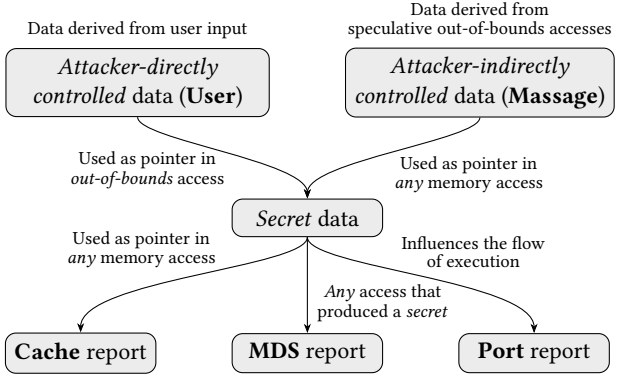
\begin{figure}[t]
\centering

\tikzstyle{flowchartnode} = [rectangle, rounded corners, minimum width=2cm, minimum height=0.5cm, text centered, draw=black, fill=gray!15, align=center, font={\small}]
\tikzstyle{flowchartinnernode} = [rectangle, rounded corners, text centered, draw=black, fill=white, align=center, font={\small}]
\tikzstyle{flowchartnote} = [font={\fontsize{7pt}{7}\selectfont}, anchor=south, align=center]

\begin{tikzpicture}[node distance=1.8cm]

\node (attacker-dir) [flowchartnode, minimum width=3.5cm] {\textit{Attacker-directly}\\\textit{controlled} data (\textbf{User})};
\node (attacker-ind) [right=1cm, flowchartnode, minimum width=3.5cm] at (attacker-dir.east) {\textit{Attacker-indirectly}\\\textit{controlled} data (\textbf{Massage})};
\node [flowchartnote] at (attacker-dir.north) {Data derived from user input};
\node [flowchartnote] at (attacker-ind.north) {Data derived from\\speculative out-of-bounds accesses};

\path (attacker-dir) -- (attacker-ind) node[midway, yshift=-1.8cm, flowchartnode] (secret) {\textit{Secret} data};

\node (mds) [below of=secret, flowchartnode] {\textbf{MDS} report};
\node (cache) [left=1cm, flowchartnode] at (mds.west) {\textbf{Cache} report};
\node (port) [right=1cm, flowchartnode] at (mds.east) {\textbf{Port} report};

\draw [-{Stealth[scale=0.75]}] (attacker-dir) .. controls ([shift={(0, -.3cm)}] attacker-dir.south) and ([shift={(0, .5cm)}] secret.north) .. 
    node[flowchartnote, xshift=-1cm, yshift=-.6cm] {Used as pointer in\\\textit{out-of-bounds} access} (secret);
\draw [-{Stealth[scale=0.75]}] (attacker-ind) .. controls ([shift={(0, -.3cm)}] attacker-ind.south) and ([shift={(0, .5cm)}] secret.north) .. 
    node[flowchartnote, xshift=1cm, yshift=-.65cm] {Used as pointer in\\\textit{any} memory access} (secret);

\draw [-{Stealth[scale=0.75]}] (secret) -- (mds) node[flowchartnote, xshift=1cm, yshift=.3cm] {\textit{Any} access that\\produced a \textit{secret}};

\draw [-{Stealth[scale=0.75]}] (secret) .. controls ([shift={(0, -0.4cm)}] secret.south) and ([shift={(0, .7cm)}] cache.north) .. 
    node[flowchartnote, xshift=-.7cm, yshift=0cm] {Used as pointer in\\\textit{any} memory access} (cache);
\draw [-{Stealth[scale=0.75]}] (secret) .. controls ([shift={(0, -0.4cm)}] secret.south) and ([shift={(0, .7cm)}] port.north) .. 
    node[flowchartnote, xshift=.7cm, yshift=0cm] {Influences the flow\\of execution} (port);

\end{tikzpicture}
\caption{Illustration of Kasper's policy on detecting Spectre-V1 gadgets (based on Fig. 3 of Kasper\cite{Kasper}).}\label{fig:kasper-policy}
\end{figure}

%% file: experiments.tex
\section{Experiments}\label{sec:experiments}

We conduct three experiments to evaluate the run-time performance and gadget finding ability of \toolname{}.
\begin{enumerate}
  \item A performance evaluation by comparing the run time of the original and instrumented binaries.
  \item A detection evaluation by fuzzing binaries with artificially inserted Spectre-V1 gadgets and examining the number of found gadgets against the ground truth.
  \item Another detection evaluation by fuzzing unmodified binaries and examining the number of found gadgets. %
\end{enumerate}
\textbf{Experimental setup.}
The experiments are conducted on an AMD EPYC 9684X (96C/192T, \SI{2.55}{\giga\hertz}) processor with \SI{768}{\giga\byte} of RAM.
We adopt the standard set of test programs used by previous studies\cite{SpecFuzz,SpecTaint}: %
\texttt{jsmn}\cite{jsmn} commit hash \texttt{18e9fe42cb}, \texttt{libyaml}\cite{libyaml} version 0.2.2, \texttt{libhtp}\cite{libhtp} version 0.5.30, \texttt{brotli}\cite{brotli} version 1.0.7 and \texttt{openssl}\cite{openssl} version 3.0.0.
The instrumented binaries are fuzzed with honggfuzz\cite{honggfuzz}, each for 24 hours and with eight threads.
We compile the fuzzing drivers
of the programs with \texttt{clang} 7.0.1 with the default compile flags specified by the programs.
In particular, \texttt{openssl} features multiple fuzzing drivers for different parts of the library, and we evaluate the \texttt{server} driver.
\textit{We do not use the source code to give \toolname{} any hint and base the instrumentation only on the binary.}

\subsection{Run-time performance}\label{ssec:exp-runtime-perf}

\input{figs/exp-runtime-overhead}

We evaluate the raw run-time performance of the instrumented programs.
Inevitably, speculation simulation brings a large run-time overhead, and instrumented programs run several magnitudes slower than their native uninstrumented counterparts.
We craft a large-sized input file for each program and measure the averaged execution time over 10 runs. \looseness=-1

We compare \toolname{} with the original SpecTaint and SpecFuzz implementations.
They utilize different heuristics to detect nested gadgets, which greatly influences run time and unfairly favors the more aggressive skipping heuristics.
For a fair comparison, we disable the nested speculation support and all heuristics for all implementations in this comparison.

Due to the lack of documentation and many hard-coded constants, we could not fully reproduce SpecTaint experiments on our test platform.
We successfully executed \texttt{jsmn} and \texttt{libyaml} with SpecTaint, but attempting to run \texttt{libhtp}, \texttt{brotli} and \texttt{openssl} crashes the emulator, and their execution times for SpecTaint are not reported.
The normalized execution times of the programs are shown in \cref{fig:exp-runtime-overhead}. \looseness=-1

Despite being a binary-only solution unable to benefit from compiler optimizations, \toolname{} demonstrates noticeably better performance: it outperforms SpecTaint, the only binary-based detector counterpart, by 22.4$\times$ in \texttt{jsmn} and 27.6$\times$ in \texttt{libyaml} respectively.
\toolname{}'s performance is also comparable with SpecFuzz, a compiler-based analyzer, in all tested applications (performance ranges between 0.5$\times$ and 2.0$\times$ of SpecFuzz).
Since \toolname{} implements a set of more complex and complete gadget policies than SpecTaint and SpecFuzz by employing binary ASan and DIFT, we attribute its superior performance to \twocopyname{}. \looseness=-1

\subsection{Detecting artificial Spectre gadgets in real-world binaries}\label{ssec:exp-fuzz-v2}

\begin{table*}[t]
  \centering
  \caption{Number of gadgets found from the evaluated real-world binary programs with artificially injected gadgets.}
  {
  \rowcolors{3}{gray90}{}
  \resizebox{\textwidth}{!}{
    {\renewcommand{\arraystretch}{1.2}
    \begin{tabular}{c<{\hspace{1pc}} c<{\hspace{1pc}} c c c c c<{\hspace{1pc}} c c c c c<{\hspace{1pc}} c c c c c<{\hspace{1pc}} c c c c c} %
      \toprule
      \multirow{2}{*}{Program} & \multirow{2}{*}{GT} & \multicolumn{5}{c}{SpecTaint (Reported\cite{SpecTaint})} & \multicolumn{5}{c}{SpecFuzz (Reported\cite{SpecTaint})} & \multicolumn{5}{c}{SpecFuzz (Reproduced)} & \multicolumn{5}{c}{\toolname{} (Proposed)} \\ & & TP & FP & FN & Precision & Recall & TP & FP & FN & Precision & Recall & TP & FP & FN & Precision & Recall & TP & FP & FN & Precision & Recall \\
      \midrule
      jsmn & 3 & 3 & 0 & 0 & 100\% & 100\% & 2 & 17 & 1 & 11\% & 67\% & 3 & 18 & 0 & 14\% & 100\% & 3 & 0 & 0 & 100\% & 100\% \\
      libyaml & 10 & 7 & 0 & 3 & 100\% & 70\% & 4 & 215 & 6 & 2\% & 40\% & 8 & 214 & 2 & 4\% & 80\% & 8 & 0 & 2 & 100\% & 80\% \\
      libhtp & 7 & 7 & 0 & 0 & 100\% & 100\% & 5 & 79 & 0 & 6\% & 71\% & 7 & 318 & 0 & 2\% & 100\% & 7 & 0 & 0 & 100\% & 100\% \\
      brotli & 13 & 12 & 0 & 1 & 100\% & 92\% & 7 & 43 & 1 & 14\% & 54\% & 13 & 297 & 0 & 4\% & 100\% & 13 & 0 & 0 & 100\% & 100\% \\
      \bottomrule
      \multicolumn{22}{c}{\cellcolor{white}\small{GT: Ground Truth; TP: True Positive; FP: False Positive; FN: False Negative. $\text{Precision} = \frac{\text{TP}}{\text{TP+FP}}$; $\text{Recall} = \frac{\text{TP}}{\text{GT}}$.}}
    \end{tabular}}\label{tab:exp-gadget-v2}
  }
}
\end{table*}

We adopt an evaluation method proposed by SpecTaint to systematically evaluate gadget detectors with solid ground truths on real-world binaries.
Specifically, we pick gadget samples from the Spectre examples\cite{Spectre15}, and artificially inject them into various positions of the programs, effectively making the programs vulnerable at these injection points.
We inject the gadgets into the same attack points used by SpecTaint in their evaluation for the results to be directly comparable.
However, as they did not make the injection points of \texttt{openssl} public, we have to drop it from this experiment. \looseness=-1

To eliminate noise from existing Spectre gadgets in the programs, we disable the taint sources and consider a variable read by the artificial gadgets as the only ``user-input'', and mark it with \textit{attacker-direct} label.
We also disable the Massage policies as they introduce the \textit{attacker-indirect} taint, which is undesired in this experiment.
Following SpecTaint's setup, we treat all gadget reports pointing to other than the artificially injected ones as false positives.

\begin{table*}[t]
  \centering
  \caption{Number of gadgets found from the evaluated real-world vanilla binary programs.}
  {
  \rowcolors{3}{gray90}{}
  \resizebox{\textwidth}{!}{
    {\renewcommand{\arraystretch}{1.2}
    \begin{tabular}{c<{\hspace{1pc}} c c<{\hspace{1pc}} c<{\hspace{1pc}} c c c c c c<{\hspace{1pc}} c c c} %
      \toprule
      \multirow{2}{*}{Program} & SpecTaint & SpecFuzz & SpecFuzz & \multicolumn{9}{c}{\toolname{} (Proposed)} \\
      & (Reported\cite{SpecTaint}) & (Reported\cite{SpecFuzz}) & (Reproduced) & User-MDS & User-Cache & User-Port & Massage-MDS & Massage-Cache & Massage-Port & Total User-* & Total Massage-* & Total *-* \\
      \midrule
      jsmn & 1 & 16 & 18 & 0 & 0 & 0 & 0 & 0 & 0 & 0 & 0 & 0 \\
      libyaml & 3 & 140 & 214 & 1 & 0 & 1 & 1 & 1 & 3 & \textbf{2} & \textbf{5} & \textbf{7} \\
      libhtp & 14 & 91 & 318 & 42 & 3 & 92 & 45 & 9 & 76 & \textbf{137} & \textbf{130} & \textbf{267} \\
      brotli & 17 & 68 & 297 & 161 & 287 & 406 & 563 & 505 & 580 & \textbf{854} & \textbf{1648} & \textbf{2502} \\
      openssl & 16 & 589 & 790 & 50 & 12 & 4 & 171 & 50 & 83 & \textbf{60} & \textbf{304} & \textbf{370} \\
      \bottomrule
    \end{tabular}}\label{tab:exp-gadget-v1}
  }
}
\end{table*}

We compare the results with SpecTaint and SpecFuzz.
For SpecFuzz, we also reproduce the experiment on our test platform to account for computation capability differences between the processors.
However, we were unable to do the same for SpecTaint; as explained in \cref{ssec:exp-runtime-perf}, we failed to reproduce it on the majority of binaries, and it is unclear how to interpret the results of SpecTaint even for the succeeded ones as its output consists of only undocumented constant numbers.
Therefore, we only report and compare with the experiment results reported in their paper.

\cref{tab:exp-gadget-v2} shows the number of detected artificial gadgets.
While SpecTaint and SpecFuzz generated more false negatives, \toolname{} achieved an almost perfect score: it only misses two gadgets in \texttt{libyaml}.
We mainly attribute the win against SpecTaint to the more comprehensive but heavy speculation heuristic, which is only practical to use because \toolname{} excels in run-time performance.
Furthermore, we manually investigated the two false negatives in \texttt{libyaml}, and found that they were inserted in modules not covered by the fuzzing driver.
Therefore, \toolname{} in fact has detected 100\% of the artificial Spectre gadgets reachable from the fuzzing drivers.

Our reproduced results of SpecFuzz in this experiment are vastly different from those reported by SpecTaint authors: SpecFuzz also detects all the gadgets in the binaries, except for the unreachable two in \texttt{libyaml}.
We believe this is because the experiments are executed on a more powerful machine.
However, SpecFuzz cannot distinguish speculative out-of-bounds accesses that the attacker-controlled and those that are unlikely exploitable, and marks all of them as gadgets.
Therefore, it generates many false positives and has low precisions.
In comparison, \toolname{} detects the same set of gadgets while generating no false positives, which is achieved because of its tainting mechanisms.

\subsection{Detecting Spectre gadgets in unmodified real-world binaries}\label{ssec:exp-fuzz-v1}

In this experiment, we fuzz the instrumented vanilla binaries \textit{without} the artificially inserted gadgets.
For \toolname{}, we categorize the gadgets by their attacker controllability and the leaking side channel.
For example, a user input controlled \textit{attacker-direct} access leaking through port contention is denoted as User-Port.

The number of detected gadgets is shown in \cref{tab:exp-gadget-v1}.
Numbers between different implementations are not directly comparable as gadget detection policies differ.
We only list them for reference. \looseness=-1
Compared with SpecTaint, \toolname{}'s User-Cache policy is the most similar but with some differences.
\toolname{} ensures that the leaked data is actually a secret by checking for out-of-bounds access, which SpecTaint is unable to do as it does not work on the program level.

Fewer gadgets are expected to be reported in \toolname{} due to fewer false positives; on the other hand, \toolname{} admittedly misses gadgets that leak via global array out-of-bounds accesses (as discussed in \cref{sssec:nahco3-asan}).
However, this is not the case we observed in \texttt{brotli} as \toolname{} detected a magnitude more User-Cache gadgets than SpecTaint.
We investigated the gadgets and found that the majority of them are protected by multiple levels of nested branches.
We therefore attribute this difference to SpecTaint's nested speculation heuristics: stopping to speculate after five tries is not enough and it misses a large number of gadgets.
This highlights the importance of an efficient gadget detector.
Within the limited time budget, we could use more extensive discovery heuristics to achieve a better tradeoff between speed and detection ability.

Our reproduced SpecFuzz results exceed the originally reported numbers, likely due to the higher computation power.
However, many of the SpecFuzz-reported gadgets are false positives due to its inability to trace the data flow.
The most relevant policy in \toolname{} with SpecFuzz is User-MDS, with the difference of \toolname{} requiring the out-of-bounds access to be attacker controllable, so definitely fewer gadgets are detected as \toolname{} eliminates the false positives.

\toolname{} also reports Spectre-V1 gadgets through various exploitation paths not implemented by other detectors, namely the User-Port and the Massage-* policies.
We detected gadgets of such types from all the tested programs except \texttt{jsmn}.
\toolname{} can be easily extended further to support other exploitation routes of Spectre and we leave this to future work. \looseness=-1

%% file: figs/exp-runtime-overhead.tex
\begin{figure}[t]
  \scalebox{0.90}{
\begin{tikzpicture}%

\begin{groupplot}[
name=overheadfig,
scaled ticks=false,
y axis line style={draw=none},
ymajorgrids=true, yminorgrids=true, minor tick num=1,
major grid style={line width=.6pt,draw=gray!60},
ybar=0pt, xtick=data, 
width=0.48\textwidth, height=6cm,
yticklabel={\pgfmathprintnumber{\tick}$\times$},
symbolic x coords={jsmn, libyaml, libhtp, brotli, openssl},
xticklabel style={yshift=1.5mm},
xtick style={draw=none}, x axis line style={line width=1pt,draw=black},
legend style={at={(1, 1)}, anchor=north east, font=\footnotesize},
legend image code/.code={\draw [#1] (0cm,-0.1cm) rectangle (0.2cm,0.15cm); },
legend cell align={left},
legend to name=overheadlegend_exp_runtime, 
legend columns=-1,
group style={
    group size=1 by 2,
    xticklabels at=edge bottom,
    vertical sep=-0.4mm,
},
]
\nextgroupplot[
  ymin=14500,ymax=20500,
  ytick distance=2000,
  x axis line style={draw=none},
  height=3.5cm,
  bar width=.3cm,
  enlarge x limits=0.15,
  y filter/.code={\pgfmathifthenelse{#1 < 14500}{NaN}{#1},},
]

\begin{scope}
\clip (axis cs: {[normalized]-1}, 14550) rectangle (axis cs: {[normalized] 5}, 20500);

\addplot[fill=black!85]
coordinates {
    (jsmn, 20196)
    (libyaml, 16984)
    (libhtp, 14500) %
    (brotli, 14500) %
    (openssl, 14500) %
};

\addplot[fill=black!55]
coordinates {
    (jsmn, 1809)
    (libyaml, 595)
    (libhtp, 347)
    (brotli, 533)
    (openssl, 0)
};

\addplot[fill=black!25]
coordinates {
    (jsmn, 900)
    (libyaml, 615)
    (libhtp, 354) 
    (brotli, 1019)
    (openssl, 0)
};
\end{scope}

\draw [decorate, decoration={snake, amplitude=0.8mm, segment length=7mm}, line width=1.5mm, black] (axis cs: {[normalized]-1}, 15000) -- (axis cs: {[normalized] 5}, 15000);
\draw [decorate, decoration={snake, amplitude=0.8mm, segment length=7mm}, line width=1mm, white] (axis cs: {[normalized]-1}, 15000) -- (axis cs: {[normalized] 5}, 15000);

\nextgroupplot[
  ymin=0,ymax=2250,
  ytick distance=1000,
  axis x line*=bottom,
  height=3cm,
  bar width=.3cm,
  enlarge x limits=0.15,
  ylabel={Normalized Run Time},
  ylabel style={xshift=1cm, yshift=0.1cm}
]

\addplot[fill=black!85]
coordinates {
    (jsmn, 20196)
    (libyaml, 16984)
    (libhtp, 0)
    (brotli, 0)
    (openssl, 0)
};

\addplot[fill=black!55]
coordinates {
    (jsmn, 1809)
    (libyaml, 595)
    (libhtp, 347)
    (brotli, 533)
    (openssl, 688)
};

\addplot[fill=black!25]
coordinates {
    (jsmn, 900)
    (libyaml, 615)
    (libhtp, 354) 
    (brotli, 1019)
    (openssl, 539)
};

\legend{SpecTaint, SpecFuzz, \toolname{} (Proposed)}
\end{groupplot}

\node () at (overheadfig.north) [below, xshift=0cm, yshift=.2cm, text=black] {\nocolorref{overheadlegend_exp_runtime}};

\end{tikzpicture}
}
\caption{Execution time of the instrumented programs with large crafted inputs, normalized to the execution time of the native versions. Lower is better. Averaged over 10 runs.}\label{fig:exp-runtime-overhead}
\end{figure}
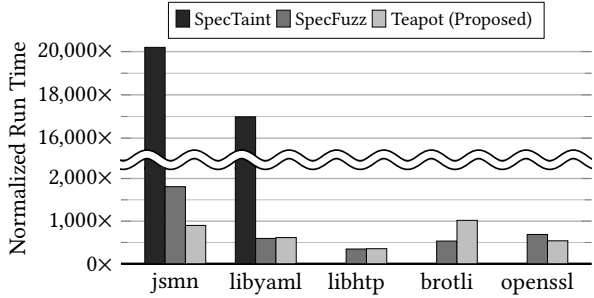

%% file: limitations.tex
\section{Limitations and Future Work}

\textbf{Correctness of rewriting.}
Due to missing source-level information such as type information and global object boundaries, many binary disassemblers are heuristic-based.
This means that incorrect disassembly results may be produced in some cases and thus binary-rewriting-based instrumentation may fail.
Unfortunately, this is a fundamental limitation of static binary rewriting.
We currently base our architecture on the state-of-the-art disassembler and rewriting platform Datalog Disassembly\cite{Ddisasm} and GTIRB\cite{GTIRB} that yield correct disassembly and instrumentation in most cases.
Since the rewriting platform is mostly orthogonal to \twocopyname{}, we might be able to enhance \toolname{} with a better alternative in the future.

\textbf{Utilization of source-level information.}
We consider \toolname{} to be a powerful supplement to compiler-based analysis even when the source code is available since it allows better consistency with the deployed binary.
However, static binary rewriting may not be the most straightforward approach when the source code is in hand.
Therefore, future work could combine \toolname{} (possibly by porting it to the compiler backend) with a traditional compiler-based detector to allow a more efficient and complete analysis.
In that case, we must use extra care to not succumb to the pitfalls discussed in \cref{ssec:motivations-compiler}.\looseness=-1

\textbf{Security trade-offs.}
Although we only implement gadget policies for detecting Spectre-V1, we could introduce support for more Spectre gadget variants; we decided to implement Spectre-V1 first as it is the most difficult to mitigate without significant overhead \cite{IntelSpectreMitigations}.
On the other hand, the limitation on AddressSanitizer granularity on stack and global objects is hard to solve without compile-time information, but this could potentially be improved with recent advancements in binary type inferencing\cite{TypeInference,Cati}.
Another fundamental trade-off shared by previous studies\cite{SpecFuzz,SpecTaint,Kasper} is that gadgets in the execution paths not covered by the fuzzing inputs cannot be detected. \looseness=-1

\textbf{Support for more instruction set architectures and operating systems.} We demonstrated the effectiveness of \toolname{} on x86-64 Linux (so that the experiment settings are in line with previous works), although realistic binary-only attack surfaces more often appear in consumer-oriented operating systems, such as Windows and macOS. \toolname{} can be extended to support other operating systems with minimum work, which is an ongoing effort. Future work could also extend \toolname{} to support different instruction sets, \eg{}AArch64 for Spectre gadget analysis on mobile platforms. \looseness=-1

%% file: related-work.tex
\section{Related Work}

\textbf{Spectre-type vulnerability detectors.}
SpecFuzz\cite{SpecFuzz}, SpecTaint\cite{SpecTaint}, Kasper\cite{Kasper} and \toolname{} leverage dynamic fuzzing to detect Spectre gadgets.
On the other hand, MSVC\cite{MSVC} and RedHat Scanner\cite{RedHat} use pattern matching for scanning Spectre gadgets, which is prone to overlook gadgets and suffer from incurring false positives and negatives\cite{SpecFuzz}.
Oo7\cite{oo7} performs static taint tracking for detecting Spectre gadgets, but still suffers from the shortcomings of static analysis and fails to cover some gadgets.
Several symbolic execution-based works
are established to discover Spectre gadgets\cite{Spectector,SpecuSym,KLEESpectre,Cats}.
Such works formalize the behavior of Spectre attacks and thus provide better security guarantees than simple pattern-matching approaches.
However, the inherent searching space explosion problem limits the usage for large-scale programs\cite{SpecFuzz}.
Recent works also utilize machine learning to detect Spectre vulnerabilities\cite{ExplainableML,HardwareRealTimeML,RealTimeDetectML}. %

\textbf{Fuzzing.}
We use honggfuzz\cite{honggfuzz} in our experiments.
Recent advancements in fuzzers such as AFL++\cite{AFL++}, AFLFast\cite{AFLFast}, VUzzer\cite{VUzzer} and Angora\cite{Angora} can potentially boost coverage and efficiency. \looseness=-1

\textbf{Binary instrumentation.}
Dynamic program analysis tools based on system emulators\cite{DECAF,QEMU} and just-in-time instrumentation\cite{Dyninst,PIN,Valgrind} inevitably suffer from a high run-time overhead.
Static reassemble disassemblers\cite{Ddisasm,RetroWrite,Uroboros} instrument the binaries on the assembly code level by creating reassemble disassembly; LLVM rewriters\cite{mctoll} further lift the binaries to LLVM IR which allows the application of LLVM compiler passes.
Such methods may suffer from the correctness of disassembly and symbolization as they try to recover compile-time information with heuristics; however, recent studies including Datalog Disassembly\cite{Ddisasm}, on which \toolname{} is based, achieve superior accuracy\cite{BinaryRewriters}.

%% file: conclusion.tex
\section{Conclusion}
\label{sec:conclusion}

We proposed \toolname{}, a Spectre gadget detector based on static binary rewriting and dynamic fuzzing, powered by \twocopyname{}.
We demonstrated the execution efficiency of \toolname{} instrumented binaries and the gadget detection effectiveness with Kasper policies.
\toolname{} proved itself to be not only a superior dynamic Spectre gadget detector for COTS binaries, but also a compelling choice for analyzing generic programs for its consistency with the deployed application. \looseness=-1
The source code of \toolname{} is available at \url{https://github.com/titech-caras/teapot}.

%% file: acknowledgements.tex
\begin{acks}

This work was supported in part by a project, JPNP23015, commissioned by the New Energy and Industrial Technology Development Organization (NEDO), and JSPS KAKENHI Grant Numbers JP22K19771 and JP24K02910.
The first author would like to express a special thanks to the legendary singer and songwriter ``Sanketsu-girl'' Sayuri, for the emotional support through her beautiful voice and songs, and may she rest in peace. 
\end{acks}

%% file: apdx-case-study.tex
\section{Case Studies}\label{sec:case-studies}\label{sec:apdx-case-studies}

We manually investigate select gadget reports of \texttt{libhtp}.
This program was selected as it is unlikely to be used in sensitive contexts, and disclosing its gadgets poses no prominent risks.
We showcase representative gadgets that can cause erroneous results (false positives and negatives) if detected with tools from previous studies, highlighting the benefit of \toolname{}. We map the vulnerabilities back to their source code for easy demonstration; actual analysis was performed on the assembly level.

\subsection{Speculative read offset manipulation}\label{ssec:case-study-1}

\begin{lstlisting}[style=my-lst-style-linenum, float=h, caption={A speculative read offset manipulation in LZMA and a subsequent cache transmitter instruction. Highlighted lines represent the exploitation route: yellow for misprediction and red for access and transmission.}, label={lst:case-study-1}]
ELzmaDummy LzmaDec_TryDummy(CLzmaDec *p) {
  // ...
  int x = p->dicPos - p->reps[0];

@\lstcolourline{yellow!15}@  // next branch mispredicted as true
@\lstcolourline{yellow!15}@  if (p->dicPos < p->reps[0])
@\lstcolourline{yellow!15}@    x += p->dicBufSize;

@\lstcolourline{red!15}@  unsigned matchByte = p->dic[x];
  unsigned offs = 0x100, symbol = 1;
  do {
    unsigned bit = offs;
    matchByte += matchByte;
    offs &= matchByte;
@\lstcolourline{red!15}@    tmp = *(prob + (offs + bit + symbol));
    // ...
  } while (symbol < 0x100);
  // ...
}
\end{lstlisting}

The User-Cache gadget shown in \cref{lst:case-study-1} is from the LZMA library, newly detected by \toolname{}.
The gadget was discovered when fuzzing \texttt{libhtp}, which internally uses the LZMA library.
The gadget allows speculative manipulation of a memory read by an attacker-controllable offset, followed by a transmitter instruction to leak the secret via the cache side channel.\looseness=-1

First, the branch on line 6 checks if the dictionary access offset is in bounds, and if it underflows, offsets it with the dictionary buffer size.
The parameters \texttt{p->dicPos} and \texttt{p->reps[0]} are not directly controllable by the attacker.
However, \texttt{p->dicBufSize} is derived from the dictionary size, which is a fundamental parameter of many compression algorithms and, in this case, carried in the metadata, which the attacker can modify.
If the branch is mispredicted, it allows an out-of-bounds access with an offset of \texttt{p->dicBufSize}, which can be manipulated by carefully constructing the input data.
The read secret is stored in the \texttt{matchByte} variable (line~9).\looseness=-1

Then, \texttt{matchByte} influences the construction of another data pointer as it is used to bit mask the \texttt{offs} variable (line~14).
The pointer dereference constitutes a transmitter instruction that leaks the secret through the cache side channel (line~15).\looseness=-1

\textbf{Challenges in detecting this vulnerability.}
This gadget may become a false positive or a false negative with compiler-based detectors.
Depending on the compiler and compile options, the \texttt{if} statement on line 6 may not generate a branch, but instead a conditional move; the gadget does not exist in the latter case since conditional moves are not speculated in modern microprocessors.
As compiler-based instrumentation enforces specific compilers and compile options, the binary code being analyzed for gadgets may not be entirely consistent with the deployed application.

\subsection{Speculative memory massage and indirectly controlled read}\label{ssec:case-study-2}

\begin{lstlisting}[style=my-lst-style-linenum, float=h, caption={A speculative memory massage and indirectly controlled read, followed by a port contention transmitter.}, label={lst:case-study-2}]
void htp_conn_destroy(htp_conn_t *conn) {
  // ...
@\lstcolourline{yellow!15}@  size_t n = list_size(conn->txs);
  for (size_t i = 0; i < n; i++) {
@\lstcolourline{red!15}@    htp_tx_t *tx = list_get(conn->txs, i);
    if (tx != NULL)  {
      htp_conn_remove_tx(tx);
      // ...
    }
  }
  // ...
}

size_t list_size(const list_t *l) {
@\lstcolourline{yellow!15}@  // next branch mispredicted as true
@\lstcolourline{yellow!15}@  if (l == NULL) return -1;
  return l->current_size;
}

void *list_get(const list_t *l, size_t idx) {
  if (l == NULL) return NULL;   
@\lstcolourline{yellow!15}@  // next two branches mispredicted
@\lstcolourline{yellow!15}@  if (idx >= l->current_size) return NULL;
@\lstcolourline{yellow!15}@  if (l->first + idx < l->max_size)
@\lstcolourline{yellow!15}@    return (void*) l->elements[l->first+idx];
  // ...
}

void htp_conn_remove_tx(htp_tx_t *tx) {
  // ... inside loop
@\lstcolourline{red!15}@  htp_tx_t *tx2 = list_get(tx->conn->txs, i);
@\lstcolourline{red!15}@  if (tx2 == tx)
    return list_replace(
      tx->conn->txs, i, NULL);
  // ...
}
\end{lstlisting}

The Massage-Port gadget in \cref{lst:case-study-2} requires three nested misspeculations to exploit.
The exploitation starts with the normal execution on line 3, where \texttt{list\_size} is called.
It is previously ensured that both \texttt{conn} and \texttt{conn->txs} are non-null.
The function \texttt{list\_size} returns the size of the list, or the error code -1 when the list is a null pointer; while the list itself is non-null, if an attacker manages to induce a misspeculation on line 16, the function returns -1.
The error code is then assigned to \texttt{n} (line 3), a \texttt{size\_t} variable.
The assignment to -1 unexpectedly makes \texttt{n} the maximum value of \texttt{size\_t}, virtually making the loop never terminating, and enabling the attacker to speculatively read data out of bounds.
Note that this assignment never happens in normal execution as \texttt{conn->txs} is ensured to be non-null.

Then, line 5 fetches the \texttt{i}-th element from the list by calling \texttt{list\_get}.
To read data out-of-bounds, the bounds checks at lines 23 and 24 must also be bypassed, requiring another two misspeculations.
This results in \texttt{tx} being a massaged pointer constructed from an out-of-bounds read, which is then passed into the \texttt{htp\_conn{\textunderscore\allowbreak}remove\_tx} function.
On line 31, some secret data is loaded into \texttt{tx2} through a series of pointer dereferencing and array accesses, which is followed by a conditional branch on line 32, completing the leak of the secret data through port contention.

\textbf{Challenges in detecting this vulnerability.} Detection of this gadget with past Spectre detector implementations is extremely challenging if not impossible.
SpecTaint does not consider exploitation through memory massaging.
While SpecFuzz may detect the initial out-of-bounds access, due to the lack of DIFT capability it cannot give further hints on the exploitation path; in this case, it spans over multiple functions and mispredictions.
Therefore, it is likely overlooked as a false positive even with a follow-up manual analysis.
Kasper does not support detecting gadgets guarded by multiple mispredictions, and its implementation does not support user-space programs.

%% file: apdx-adae.tex
\newcommand\dockerteapot{\colorbox{myverylightblue}{\texttt{teapot}}\,}
\newcommand\dockerspecfuzz{\colorbox{myyellow}{\texttt{specfuzz}}\,}

\section{Artifact Appendix}

\subsection{Abstract}
The artifact contains Teapot, the first Spectre gadget detector for COTS binaries with comparable performance to compiler-based detectors.
Teapot is composed of a static binary rewriter and a runtime support library.
The binary programs of interest are first statically instrumented by Teapot, then dynamically fuzzed to identify Spectre gadgets.

The artifact is a self-contained archival file with all necessary data sets, programs, dependencies, and scripts to reproduce the experiments in this work.
A related work, SpecFuzz\cite{SpecFuzz}, is also included as the comparison baseline, and its results can be reproduced.

\subsection{Artifact check-list (meta-information)}

{\small
\begin{itemize}
  \item {\textbf{Program:} Teapot static rewriter, runtime library and modified \texttt{honggfuzz}. SpecFuzz is also included as baseline.}
  \item {\textbf{Compilation:} clang 10.0.0 for the runtime library; clang 7.0.1 for the test programs.}
  \item {\textbf{Data set:} 5 real-world programs (\texttt{jsmn}, \texttt{yaml}, \texttt{libhtp}, \texttt{brotli}, \texttt{openssl}).}
  \item {\textbf{Run-time environment:} x86-64 Linux.}
  \item {\textbf{Hardware:} x86-64 platform preferably with 8 or more cores.}
  \item {\textbf{Execution:} scripts are provided to compile, instrument and benchmark the test programs.}
  \item {\textbf{Metrics:} execution time and the number of gadgets found.}
  \item {\textbf{Output:} raw data that corresponds to \cref{fig:exp-runtime-overhead} and \cref{tab:exp-gadget-v1}.}
  \item {\textbf{Experiments:} (1) execution time comparison between instrumented and vanilla binaries (\cref{fig:exp-runtime-overhead}); (2) number of gadgets found in real-world binaries (\cref{tab:exp-gadget-v1}).}
  \item {\textbf{How much disk space required (approximately)?:} a few gigabytes of free space.}
  \item {\textbf{How much time is needed to prepare workflow (approximately)?:} 1 to 2 hours except \texttt{openssl}; \texttt{openssl} requires a few hours of preparation.}
  \item {\textbf{How much time is needed to complete experiments (approximately)?:} 1 day per experiment (10 in total, can run in parallel). However, approximate results can be achieved within 3 hours to save time (see \cref{sssec:adae-reduce-fuzzing-time}).}
  \item {\textbf{Publicly available?:} Yes. See \cref{sec:conclusion}.}
  \item {\textbf{Code licenses (if publicly available)?:} GPLv3}
  \item {\textbf{Archived (provide DOI)?:} \href{https://doi.org/10.5281/zenodo.14507732}{10.5281/zenodo.14507732}} \cite{TeapotArtifact}
\end{itemize}}

\subsection{Description}

\subsubsection{How delivered}
The artifact is available as a self-contained archive at \url{https://doi.org/10.5281/zenodo.14507732}, or alternatively from \url{https://github.com/titech-caras/teapot-artifact}.

\subsubsection{Hardware dependencies}
The artifact must be executed on an x86-64 platform, preferably with 8 or more cores (the artifact assumes 8 execution threads are available). 
Note that rewriting the \texttt{openssl} binary requires significantly more RAM (40+ GB) than other experiments. 
The rest of the experiments can run on a system with limited RAM.

\subsubsection{Software dependencies}

We recommend setting up the experiment using the provided \texttt{Dockerfile}s, as the fairly complex dependencies will be automatically configured. 
For manual setup, detailed dependency requirements are discussed in \texttt{teapot/README.md}.

\subsubsection{Data sets}
A set of 5 programs are evaluated: \texttt{jsmn}\cite{jsmn} commit hash \texttt{18e9fe42cb}, \texttt{yaml}\cite{libyaml} version 0.2.2, \texttt{libhtp}\cite{libhtp} version 0.5.30, \texttt{brotli}\cite{brotli} version 1.0.7 and \texttt{openssl}\cite{openssl} version 3.0.0.

The source code of these programs is contained in the artifact and need not be prepared manually.
Compilation and instrumentation of these programs are required, and scripts are provided to assist the process.

\subsection{Installation}

Two docker containers are used throughout the experiments. First, we build the images with the provided \texttt{Dockerfile}s.

\begin{lstlisting}[style=my-lst-style-bg]
$ cd /path/to/teapot-artifact
$ docker build -t teapot_img teapot/
$ docker build -t specfuzz_img specfuzz/
\end{lstlisting}

Next, we create the two containers from the images, and name them \dockerteapot and \dockerspecfuzz correspondingly. 
We mount the artifact folder at \texttt{/workspace} and launch two terminals for conducting experiments.

\begin{lstlisting}[style=my-lst-style-bg]
$ docker run --name teapot -it -v /path/to/teapot-artifact:/workspace teapot_img bash
$ docker run --name specfuzz -it -v /path/to/teapot-artifact:/workspace specfuzz_img bash
\end{lstlisting}

It may also be desirable to add \texttt{----user \$(id --u):\$(id --g)} to avoid file permission problems.

\subsection{Experiment workflow}

\textbf{Important:} we demonstrate the workflow with \texttt{libhtp}. Replace \texttt{libhtp} with \texttt{\{brotli,openssl,yaml,jsmn\}} for experiments on the remaining binaries.

\subsubsection{Compile and instrument the test programs}
First, we compile the programs to be tested.
This step generally takes a few minutes to half an hour for each program, but can be extra lengthy for \texttt{openssl}.
We use the compiler toolkit from the \dockerspecfuzz container.
Execute the following command in \dockerspecfuzz:

\begin{lstlisting}[style=my-lst-style-bg]
$ scripts/compile/libhtp.sh
\end{lstlisting}

The script generates two files: \texttt{binaries/original/lib{\allowbreak}htp}, the vanilla uninstrumented binary; \texttt{binaries/specfuzz{\allowbreak}/libhtp}, the SpecFuzz-instrumented binary (used for both runtime performance evaluation and fuzzing).

Instrumentation with Teapot requires running one extra script. 
Note that this step takes extra RAM for \texttt{openssl}.
Execute the following command in \dockerteapot:

\begin{lstlisting}[style=my-lst-style-bg]
$ scripts/instrument_teapot.sh libhtp
\end{lstlisting}

The script generates two files: \texttt{binaries/teapot/libhtp}, the Teapot-instrumented binary with nested speculation (used for fuzzing), and \texttt{binaries/teapot\_nonest/libhtp}, the Teapot-instrumented binary without nested speculation (used for runtime performance evaluation). 

\subsubsection{Run-time performance of instrumented programs}

Executing the following script in \dockerteapot automatic- ally evaluates the run-time performance of the uninstrumented binary, the SpecFuzz- and the Teapot-instrumented binary\footnote{SpecFuzz binaries are statically linked and can be executed in any environment. Teapot binaries have dynamically linked dependencies, which is why this experiment is conducted in \dockerteapot.}. This requires a few minutes to half an hour for each program.

\begin{lstlisting}[style=my-lst-style-bg]
$ scripts/exp/eval_runtime_perf.sh libhtp
\end{lstlisting}

The script generates \texttt{results/runtime/libhtp.json} which contains the raw results for generating \cref{fig:exp-runtime-overhead}.

\subsubsection{Fuzzing the programs for real-world Spectre gadgets}
\label{sssec:adae-fuzzing}

Executing the following script in \dockerteapot starts fuzzing the Teapot-instrumented binary program with \texttt{hongg\-fuzz} supplied in the container.
The last parameter is the time budget for fuzzing in seconds; for 1 day (following the experiment setup in \cref{ssec:exp-fuzz-v1}) enter \texttt{86400}.

\begin{lstlisting}[style=my-lst-style-bg]
$ scripts/exp/fuzz_teapot.sh libhtp 86400
\end{lstlisting}

The script generates \texttt{results/fuzz/teapot/libhtp.{\allowbreak}json}, which contains the Teapot part of raw results later used for generating \cref{tab:exp-gadget-v1}.

As a comparison baseline, executing the following script in \dockerspecfuzz starts fuzzing the SpecFuzz-instrumented binary program with SpecFuzz-specific \texttt{honggfuzz}.

\begin{lstlisting}[style=my-lst-style-bg]
$ scripts/exp/fuzz_specfuzz.sh libhtp 86400
\end{lstlisting}

The script generates \texttt{results/fuzz/specfuzz/libhtp.{\allowbreak}json}, which contains the SpecFuzz part of raw results later used for generating \cref{tab:exp-gadget-v1}.

\subsection{Evaluation and expected result}
Once the result collection for all binaries is complete, execute the analysis scripts in \dockerteapot to generate aggregated data. 

\begin{lstlisting}[style=my-lst-style-bg]
$ scripts/analysis/runtime.py
$ scripts/analysis/fuzz.py
\end{lstlisting}

The scripts generate \texttt{results/runtime\_aggregated.{\allowbreak}csv} and \texttt{results/fuzz\_aggregated.csv}. 
By pasting their contents into a provided Microsoft Excel spreadsheet file \texttt{results/{\allowbreak}aggregated.xlsx}, a figure similar to \cref{fig:exp-runtime-overhead} and a table corresponding to \cref{tab:exp-gadget-v1} are generated.

\subsection{Experiment customization}

\subsubsection{Parallelization of installation and setup}
The installation scripts default to use 8 threads for compilation.
Compilation of LLVM when building \dockerspecfuzz image, and the compilation of \texttt{openssl} may be slow under this setting.

Modifying the thread parameters in \texttt{specfuzz/install/{\allowbreak}llvm.sh} and \texttt{scripts/compile/common.sh} would allow more threads to be used, providing noticeable speedups.

\subsubsection{Parallelization of experiments}
The fuzzing experiment described in \cref{sssec:adae-fuzzing} is very time-consuming. 
The 10 tasks in the experiment, however, can be executed in parallel on different cores/machines.
In practice, we executed the experiments on one AMD EPYC 9684X and pinned each task (8 threads) onto one CCD chiplet (containing exactly 8 cores).

If the experiments span across multiple machines, the result files (\texttt{results/fuzz/\{teapot,specfuzz\}/*.json}) must be collected before executing the analysis scripts.

\subsubsection{Reducing fuzzing time} 
\label{sssec:adae-reduce-fuzzing-time}
While we fuzz each binary for one day, we find that many gadgets can already be detected in as little as three hours. 
Therefore, an approximation can be achieved by setting the time limit of each task in \cref{sssec:adae-fuzzing} to 3 hours, if time is limited.
In this case, change the time budget parameter to \texttt{10800}.

Alternatively, the thread count (default 8) for fuzzing can also be changed by modifying a parameter in \texttt{scripts/exp/{\allowbreak}fuzz\_\{teapot,specfuzz\}.sh}.

\subsection{Notes}

\subsubsection{Troubleshooting}
We expect the artifact scripts to execute without errors (warnings may be present, but most can be safely ignored).

Nevertheless, we collected common issues that may occur with Teapot (typically when executed outside provided containers and/or without using the scripts), described in \texttt{teapot/TROUBLESHOOTING.md} in further detail.

\subsubsection{Limitations}

The following experiments are not reproduced in this artifact:

\begin{itemize}
    \item The experiment on artificial gadgets (\cref{tab:exp-gadget-v2}); while automatically generating the raw data is possible, the interpretation is non-trivial. Corresponding the gadgets found to the source code is difficult, but required for differentiating true positives, false positives, and false negatives. This is done by manual analysis, which requires significant time and reverse engineering experience. Since the raw data alone tells little of the story, we omit this experiment in the scripts.
    \item SpecTaint results in \cref{fig:exp-runtime-overhead}; reproducing their experiments requires significant manual effort, and its execution is unreliable and often has to be retried multiple times due to random crashes. This is ultimately because of their lack of documentation, as discussed in \cref{ssec:exp-runtime-perf}.
\end{itemize}

%% file: main.bbl

\begin{thebibliography}{49}


\ifx \showCODEN    \undefined \def \showCODEN     #1{\unskip}     \fi
\ifx \showDOI      \undefined \def \showDOI       #1{#1}\fi
\ifx \showISBNx    \undefined \def \showISBNx     #1{\unskip}     \fi
\ifx \showISBNxiii \undefined \def \showISBNxiii  #1{\unskip}     \fi
\ifx \showISSN     \undefined \def \showISSN      #1{\unskip}     \fi
\ifx \showLCCN     \undefined \def \showLCCN      #1{\unskip}     \fi
\ifx \shownote     \undefined \def \shownote      #1{#1}          \fi
\ifx \showarticletitle \undefined \def \showarticletitle #1{#1}   \fi
\ifx \showURL      \undefined \def \showURL       {\relax}        \fi
\providecommand\bibfield[2]{#2}
\providecommand\bibinfo[2]{#2}
\providecommand\natexlab[1]{#1}
\providecommand\showeprint[2][]{arXiv:#2}

\bibitem[Ahmad(2020)]%
        {RealTimeDetectML}
\bibfield{author}{\bibinfo{person}{Bilal~Ali Ahmad}.} \bibinfo{year}{2020}\natexlab{}.
\newblock \showarticletitle{Real time Detection of Spectre and Meltdown Attacks Using Machine Learning}.
\newblock \bibinfo{journal}{\emph{arXiv preprint arXiv:2006.01442}} (\bibinfo{year}{2020}).
\newblock


\bibitem[Bellard(2005)]%
        {QEMU}
\bibfield{author}{\bibinfo{person}{Fabrice Bellard}.} \bibinfo{year}{2005}\natexlab{}.
\newblock \showarticletitle{{QEMU}, A Fast and Portable Dynamic Translator}. In \bibinfo{booktitle}{\emph{Proceedings of the 2005 {USENIX} Annual Technical Conference}} \emph{(\bibinfo{series}{USENIX ATC~'05})}.
\newblock


\bibitem[Bernat and Miller(2011)]%
        {Dyninst}
\bibfield{author}{\bibinfo{person}{Andrew~R. Bernat} {and} \bibinfo{person}{Barton~P. Miller}.} \bibinfo{year}{2011}\natexlab{}.
\newblock \showarticletitle{Anywhere, Any-Time Binary Instrumentation}. In \bibinfo{booktitle}{\emph{Proceedings of the 10th {ACM} {SIGPLAN-SIGSOFT} Workshop on Program Analysis for Software Tools}} \emph{(\bibinfo{series}{PASTE~'11})}.
\newblock


\bibitem[B{\"o}hme et~al\mbox{.}(2016)]%
        {AFLFast}
\bibfield{author}{\bibinfo{person}{Marcel B{\"o}hme}, \bibinfo{person}{Van-Thuan Pham}, {and} \bibinfo{person}{Abhik Roychoudhury}.} \bibinfo{year}{2016}\natexlab{}.
\newblock \showarticletitle{Coverage-Based Greybox Fuzzing as {Markov} Chain}. In \bibinfo{booktitle}{\emph{Proceedings of the 2016 {ACM} {SIGSAC} Conference on Computer and Communications Security}} \emph{(\bibinfo{series}{CCS~'16})}.
\newblock


\bibitem[Caballero and Lin(2016)]%
        {TypeInference}
\bibfield{author}{\bibinfo{person}{Juan Caballero} {and} \bibinfo{person}{Zhiqiang Lin}.} \bibinfo{year}{2016}\natexlab{}.
\newblock \showarticletitle{Type inference on executables}.
\newblock \bibinfo{journal}{\emph{ACM Computing Surveys (CSUR)}} \bibinfo{volume}{48}, \bibinfo{number}{4} (\bibinfo{year}{2016}), \bibinfo{pages}{1--35}.
\newblock


\bibitem[Canella et~al\mbox{.}(2019)]%
        {Fallout}
\bibfield{author}{\bibinfo{person}{Claudio Canella}, \bibinfo{person}{Daniel Genkin}, \bibinfo{person}{Lukas Giner}, \bibinfo{person}{Daniel Gruss}, \bibinfo{person}{Moritz Lipp}, \bibinfo{person}{Marina Minkin}, \bibinfo{person}{Daniel Moghimi}, \bibinfo{person}{Frank Piessens}, \bibinfo{person}{Michael Schwarz}, \bibinfo{person}{Berk Sunar}, \bibinfo{person}{Jo Van~Bulck}, {and} \bibinfo{person}{Yuval Yarom}.} \bibinfo{year}{2019}\natexlab{}.
\newblock \showarticletitle{Fallout: Leaking Data on {Meltdown}-resistant {CPUs}}. In \bibinfo{booktitle}{\emph{Proceedings of the {ACM} {SIGSAC} Conference on Computer and Communications Security}} \emph{(\bibinfo{series}{CCS~'19})}.
\newblock


\bibitem[Chen et~al\mbox{.}(2020)]%
        {Cati}
\bibfield{author}{\bibinfo{person}{Ligeng Chen}, \bibinfo{person}{Zhongling He}, {and} \bibinfo{person}{Bing Mao}.} \bibinfo{year}{2020}\natexlab{}.
\newblock \showarticletitle{Cati: Context-assisted Type Inference from Stripped Binaries}. In \bibinfo{booktitle}{\emph{Proceedings of the 2020 50th Annual IEEE/IFIP International Conference on Dependable Systems and Networks}} \emph{(\bibinfo{series}{DSN~'20})}.
\newblock


\bibitem[Chen and Chen(2018)]%
        {Angora}
\bibfield{author}{\bibinfo{person}{Peng Chen} {and} \bibinfo{person}{Hao Chen}.} \bibinfo{year}{2018}\natexlab{}.
\newblock \showarticletitle{Angora: Efficient Fuzzing by Principled Search}. In \bibinfo{booktitle}{\emph{Proceedings of the 2018 {IEEE} Symposium on Security and Privacy}} \emph{(\bibinfo{series}{IEEE S\&P~'18})}.
\newblock


\bibitem[Clifton(2018)]%
        {RedHat}
\bibfield{author}{\bibinfo{person}{Nick Clifton}.} \bibinfo{year}{2018}\natexlab{}.
\newblock \bibinfo{title}{{SPECTRE} Variant 1 Scanning Tool}.
\newblock
\newblock
\urldef\tempurl%
\url{https://access.redhat.com/blogs/766093/posts/3510331}
\showURL{%
Retrieved 2023-11-29 from \tempurl}


\bibitem[Dinesh et~al\mbox{.}(2020)]%
        {RetroWrite}
\bibfield{author}{\bibinfo{person}{Sushant Dinesh}, \bibinfo{person}{Nathan Burow}, \bibinfo{person}{Dongyan Xu}, {and} \bibinfo{person}{Mathias Payer}.} \bibinfo{year}{2020}\natexlab{}.
\newblock \showarticletitle{Retrowrite: Statically Instrumenting {COTS} Binaries for Fuzzing and Sanitization}. In \bibinfo{booktitle}{\emph{Proceedings of the 2020 {IEEE} Symposium on Security and Privacy}} \emph{(\bibinfo{series}{IEEE S\&P~'20})}.
\newblock


\bibitem[Fioraldi et~al\mbox{.}(2020)]%
        {AFL++}
\bibfield{author}{\bibinfo{person}{Andrea Fioraldi}, \bibinfo{person}{Dominik Maier}, \bibinfo{person}{Heiko Ei{\ss}feldt}, {and} \bibinfo{person}{Marc Heuse}.} \bibinfo{year}{2020}\natexlab{}.
\newblock \showarticletitle{{AFL}++: Combining Incremental Steps of Fuzzing Research}. In \bibinfo{booktitle}{\emph{Proceedings of the 14th {USENIX} Workshop on Offensive Technologies}} \emph{(\bibinfo{series}{WOOT~'20})}.
\newblock


\bibitem[Flores-Montoya and Schulte(2020)]%
        {Ddisasm}
\bibfield{author}{\bibinfo{person}{Antonio Flores-Montoya} {and} \bibinfo{person}{Eric Schulte}.} \bibinfo{year}{2020}\natexlab{}.
\newblock \showarticletitle{Datalog Disassembly}. In \bibinfo{booktitle}{\emph{Proceedings of the 29th {USENIX} Security Symposium}} \emph{(\bibinfo{series}{USENIX Security~'20})}.
\newblock


\bibitem[Foundation(2023)]%
        {libhtp}
\bibfield{author}{\bibinfo{person}{Open Information~Security Foundation}.} \bibinfo{year}{2023}\natexlab{}.
\newblock \bibinfo{title}{libhtp: {LibHTP} is A Security-Aware Parser for the {HTTP} Protocol and the Related Bits and Pieces}.
\newblock
\newblock
\urldef\tempurl%
\url{https://github.com/OISF/libhtp}
\showURL{%
Retrieved 2023-11-29 from \tempurl}


\bibitem[{Google}(2023a)]%
        {brotli}
\bibfield{author}{\bibinfo{person}{{Google}}.} \bibinfo{year}{2023}\natexlab{a}.
\newblock \bibinfo{title}{brotli: {Brotli} Compression Format}.
\newblock
\newblock
\urldef\tempurl%
\url{https://github.com/google/brotli}
\showURL{%
Retrieved 2023-11-29 from \tempurl}


\bibitem[{Google}(2023b)]%
        {honggfuzz}
\bibfield{author}{\bibinfo{person}{{Google}}.} \bibinfo{year}{2023}\natexlab{b}.
\newblock \bibinfo{title}{Honggfuzz: Security Oriented Software Fuzzer}.
\newblock
\newblock
\urldef\tempurl%
\url{https://github.com/google/honggfuzz}
\showURL{%
Retrieved 2023-11-29 from \tempurl}


\bibitem[Guarnieri et~al\mbox{.}(2020)]%
        {Spectector}
\bibfield{author}{\bibinfo{person}{Marco Guarnieri}, \bibinfo{person}{Boris K{\"o}pf}, \bibinfo{person}{Jos{\'e}~F Morales}, \bibinfo{person}{Jan Reineke}, {and} \bibinfo{person}{Andr{\'e}s S{\'a}nchez}.} \bibinfo{year}{2020}\natexlab{}.
\newblock \showarticletitle{Spectector: Principled Detection of Speculative Information Flows}. In \bibinfo{booktitle}{\emph{Proceedings of the 2020 {IEEE} Symposium on Security and Privacy}} \emph{(\bibinfo{series}{IEEE S\&P~'20})}.
\newblock


\bibitem[Guo et~al\mbox{.}(2020)]%
        {SpecuSym}
\bibfield{author}{\bibinfo{person}{Shengjian Guo}, \bibinfo{person}{Yueqi Chen}, \bibinfo{person}{Peng Li}, \bibinfo{person}{Yueqiang Cheng}, \bibinfo{person}{Huibo Wang}, \bibinfo{person}{Meng Wu}, {and} \bibinfo{person}{Zhiqiang Zuo}.} \bibinfo{year}{2020}\natexlab{}.
\newblock \showarticletitle{{SpecuSym}: Speculative Symbolic Execution for Cache Timing Leak Detection}. In \bibinfo{booktitle}{\emph{Proceedings of the ACM/IEEE 42nd International Conference on Software Engineering}} \emph{(\bibinfo{series}{ICSE~'20})}.
\newblock


\bibitem[Henderson et~al\mbox{.}(2014)]%
        {DECAF}
\bibfield{author}{\bibinfo{person}{Andrew Henderson}, \bibinfo{person}{Aravind Prakash}, \bibinfo{person}{Lok~Kwong Yan}, \bibinfo{person}{Xunchao Hu}, \bibinfo{person}{Xujiewen Wang}, \bibinfo{person}{Rundong Zhou}, {and} \bibinfo{person}{Heng Yin}.} \bibinfo{year}{2014}\natexlab{}.
\newblock \showarticletitle{Make It Work, Make It Right, Make It Fast: Building a Platform-Neutral Whole-System Dynamic Binary Analysis Platform}. In \bibinfo{booktitle}{\emph{Proceedings of the 2014 International Symposium on Software Testing and Analysis}} \emph{(\bibinfo{series}{ISSTA~'14})}.
\newblock


\bibitem[Hennessy and Patterson(2017)]%
        {HennessyPatterson}
\bibfield{author}{\bibinfo{person}{John~L Hennessy} {and} \bibinfo{person}{David~A Patterson}.} \bibinfo{year}{2017}\natexlab{}.
\newblock \bibinfo{booktitle}{\emph{Computer Architecture: A Quantitative Approach, Sixth Edition}}.
\newblock \bibinfo{publisher}{Morgan Kaufmann}.
\newblock


\bibitem[Intel(2018)]%
        {IntelSpectreMitigations}
\bibfield{author}{\bibinfo{person}{Intel}.} \bibinfo{year}{2018}\natexlab{}.
\newblock \bibinfo{title}{Speculative Execution Side Channel Mitigations}.
\newblock
\newblock
\urldef\tempurl%
\url{https://www.intel.com/content/dam/develop/external/us/en/documents/336996-speculative-execution-side-channel-mitigations.pdf}
\showURL{%
Retrieved 2024-07-08 from \tempurl}


\bibitem[Johannesmeyer et~al\mbox{.}(2022)]%
        {Kasper}
\bibfield{author}{\bibinfo{person}{Brian Johannesmeyer}, \bibinfo{person}{Jakob Koschel}, \bibinfo{person}{Kaveh Razavi}, \bibinfo{person}{Herbert Bos}, {and} \bibinfo{person}{Cristiano Giuffrida}.} \bibinfo{year}{2022}\natexlab{}.
\newblock \showarticletitle{Kasper: Scanning for Generalized Transient Execution Gadgets in the {Linux} Kernel}. In \bibinfo{booktitle}{\emph{Proceedings of the 29th Network and Distributed System Security Symposium}} \emph{(\bibinfo{series}{NDSS~'22})}.
\newblock


\bibitem[Kocher(2018)]%
        {Spectre15}
\bibfield{author}{\bibinfo{person}{Paul Kocher}.} \bibinfo{year}{2018}\natexlab{}.
\newblock \bibinfo{title}{{Spectre} Mitigations in {Microsoft's} {C/C++} Compiler}.
\newblock
\newblock
\urldef\tempurl%
\url{https://www.paulkocher.com/doc/MicrosoftCompilerSpectreMitigation.html}
\showURL{%
Retrieved 2023-11-29 from \tempurl}


\bibitem[Kocher et~al\mbox{.}(2019)]%
        {SpectrePHTBTB}
\bibfield{author}{\bibinfo{person}{Paul Kocher}, \bibinfo{person}{Jann Horn}, \bibinfo{person}{Anders Fogh}, \bibinfo{person}{Daniel Genkin}, \bibinfo{person}{Daniel Gruss}, \bibinfo{person}{Werner Haas}, \bibinfo{person}{Mike Hamburg}, \bibinfo{person}{Moritz Lipp}, \bibinfo{person}{Stefan Mangard}, \bibinfo{person}{Thomas Prescher}, \bibinfo{person}{Michael Schwarz}, {and} \bibinfo{person}{Yuval Yarom}.} \bibinfo{year}{2019}\natexlab{}.
\newblock \showarticletitle{Spectre Attacks: Exploiting Speculative Execution}. In \bibinfo{booktitle}{\emph{Proceedings of the 40th {IEEE} Symposium on Security and Privacy}} \emph{(\bibinfo{series}{IEEE S\&P~'19})}.
\newblock


\bibitem[Lin et~al\mbox{.}(2024)]%
        {TeapotArtifact}
\bibfield{author}{\bibinfo{person}{Fangzheng Lin}, \bibinfo{person}{Zhongfa Wang}, {and} \bibinfo{person}{Hiroshi Sasaki}.} \bibinfo{year}{2024}\natexlab{}.
\newblock \bibinfo{booktitle}{\emph{Artifact of the paper "Teapot: Efficiently Uncovering Spectre Gadgets in COTS Binaries"}}.
\newblock
\urldef\tempurl%
\url{https://doi.org/10.5281/zenodo.14507732}
\showDOI{\tempurl}


\bibitem[{LLVM}(2023a)]%
        {DFSan}
\bibfield{author}{\bibinfo{person}{{LLVM}}.} \bibinfo{year}{2023}\natexlab{a}.
\newblock \bibinfo{title}{{DataFlowSanitizer}}.
\newblock
\newblock
\urldef\tempurl%
\url{https://clang.llvm.org/docs/DataFlowSanitizer.html}
\showURL{%
Retrieved 2023-11-29 from \tempurl}


\bibitem[{LLVM}(2023b)]%
        {SanCov}
\bibfield{author}{\bibinfo{person}{{LLVM}}.} \bibinfo{year}{2023}\natexlab{b}.
\newblock \bibinfo{title}{{SanitizerCoverage}}.
\newblock
\newblock
\urldef\tempurl%
\url{https://clang.llvm.org/docs/SanitizerCoverage.html}
\showURL{%
Retrieved 2023-11-29 from \tempurl}


\bibitem[Luk et~al\mbox{.}(2005)]%
        {PIN}
\bibfield{author}{\bibinfo{person}{Chi-Keung Luk}, \bibinfo{person}{Robert Cohn}, \bibinfo{person}{Robert Muth}, \bibinfo{person}{Harish Patil}, \bibinfo{person}{Artur Klauser}, \bibinfo{person}{Geoff Lowney}, \bibinfo{person}{Steven Wallace}, \bibinfo{person}{Vijay~Janapa Reddi}, {and} \bibinfo{person}{Kim Hazelwood}.} \bibinfo{year}{2005}\natexlab{}.
\newblock \showarticletitle{Pin: Building Customized Program Analysis Tools with Dynamic Instrumentation}. In \bibinfo{booktitle}{\emph{Proceedings of the 26th {ACM} {SIGPLAN} Conference on Programming Language Design and Implementation}} \emph{(\bibinfo{series}{PLDI~'05})}.
\newblock


\bibitem[Microsoft(2021)]%
        {MSVC}
\bibfield{author}{\bibinfo{person}{Microsoft}.} \bibinfo{year}{2021}\natexlab{}.
\newblock \bibinfo{title}{{MSVC} Compiler Reference: {/Qspectre}}.
\newblock
\newblock
\urldef\tempurl%
\url{https://learn.microsoft.com/en-us/cpp/build/reference/qspectre?view=msvc-170}
\showURL{%
Retrieved 2023-11-29 from \tempurl}


\bibitem[Nethercote and Seward(2007)]%
        {Valgrind}
\bibfield{author}{\bibinfo{person}{Nicholas Nethercote} {and} \bibinfo{person}{Julian Seward}.} \bibinfo{year}{2007}\natexlab{}.
\newblock \showarticletitle{Valgrind: A Framework for Heavyweight Dynamic Binary Instrumentation}. In \bibinfo{booktitle}{\emph{Proceedings of the 28th {ACM} {SIGPLAN} Conference on Programming Language Design and Implementation}} \emph{(\bibinfo{series}{PLDI~'07})}.
\newblock


\bibitem[Oleksenko et~al\mbox{.}(2020)]%
        {SpecFuzz}
\bibfield{author}{\bibinfo{person}{Oleksii Oleksenko}, \bibinfo{person}{Bohdan Trach}, \bibinfo{person}{Mark Silberstein}, {and} \bibinfo{person}{Christof Fetzer}.} \bibinfo{year}{2020}\natexlab{}.
\newblock \showarticletitle{{SpecFuzz}: Bringing {Spectre}-Type Vulnerabilities to the Surface}. In \bibinfo{booktitle}{\emph{Proceedings of the 29th {USENIX} Security Symposium}} \emph{(\bibinfo{series}{USENIX Security~'20})}.
\newblock


\bibitem[{OpenSSL}(2023)]%
        {openssl}
\bibfield{author}{\bibinfo{person}{{OpenSSL}}.} \bibinfo{year}{2023}\natexlab{}.
\newblock \bibinfo{title}{openssl: {TLS/SSL} and Crypto Library}.
\newblock
\newblock
\urldef\tempurl%
\url{https://github.com/openssl/openssl}
\showURL{%
Retrieved 2023-11-29 from \tempurl}


\bibitem[Osvik et~al\mbox{.}(2006)]%
        {EVICT+TIME_PRIME+PROBE}
\bibfield{author}{\bibinfo{person}{Dag~Arne Osvik}, \bibinfo{person}{Adi Shamir}, {and} \bibinfo{person}{Eran Tromer}.} \bibinfo{year}{2006}\natexlab{}.
\newblock \showarticletitle{Cache Attacks and Countermeasures: The Case of {AES}}. In \bibinfo{booktitle}{\emph{Proceedings of the Cryptographers' Track at the RSA Conference}} \emph{(\bibinfo{series}{CT-RSA~'06})}.
\newblock


\bibitem[Pan and Mishra(2021)]%
        {ExplainableML}
\bibfield{author}{\bibinfo{person}{Zhixin Pan} {and} \bibinfo{person}{Prabhat Mishra}.} \bibinfo{year}{2021}\natexlab{}.
\newblock \showarticletitle{Automated Detection of {Spectre} and {Meltdown} Attacks Using Explainable Machine Learning}. In \bibinfo{booktitle}{\emph{Proceedings of the 2021 IEEE International Symposium on Hardware Oriented Security and Trust}} \emph{(\bibinfo{series}{HOST~'21})}.
\newblock


\bibitem[Ponce-de Le{\'o}n and Kinder(2022)]%
        {Cats}
\bibfield{author}{\bibinfo{person}{Hern{\'a}n Ponce-de Le{\'o}n} {and} \bibinfo{person}{Johannes Kinder}.} \bibinfo{year}{2022}\natexlab{}.
\newblock \showarticletitle{Cats vs. {Spectre}: An Axiomatic Approach to Modeling Speculative Execution Attacks}. In \bibinfo{booktitle}{\emph{Proceedings of the 2022 {IEEE} Symposium on Security and Privacy}} \emph{(\bibinfo{series}{IEEE S\&P~'22})}.
\newblock


\bibitem[Project(2021)]%
        {libyaml}
\bibfield{author}{\bibinfo{person}{The~{YAML} Project}.} \bibinfo{year}{2021}\natexlab{}.
\newblock \bibinfo{title}{{LibYAML} - A {C} library for Parsing and Emitting {YAML}}.
\newblock
\newblock
\urldef\tempurl%
\url{https://github.com/yaml/libyaml}
\showURL{%
Retrieved 2023-11-29 from \tempurl}


\bibitem[Qi et~al\mbox{.}(2021)]%
        {SpecTaint}
\bibfield{author}{\bibinfo{person}{Zhenxiao Qi}, \bibinfo{person}{Qian Feng}, \bibinfo{person}{Yueqiang Cheng}, \bibinfo{person}{Mengjia Yan}, \bibinfo{person}{Peng Li}, \bibinfo{person}{Heng Yin}, {and} \bibinfo{person}{Tao Wei}.} \bibinfo{year}{2021}\natexlab{}.
\newblock \showarticletitle{{SpecTaint}: Speculative Taint Analysis for Discovering {Spectre} Gadgets}. In \bibinfo{booktitle}{\emph{Proceedings of the 28th Network and Distributed System Security Symposium}} \emph{(\bibinfo{series}{NDSS~'21})}.
\newblock


\bibitem[Rawat et~al\mbox{.}(2017)]%
        {VUzzer}
\bibfield{author}{\bibinfo{person}{Sanjay Rawat}, \bibinfo{person}{Vivek Jain}, \bibinfo{person}{Ashish Kumar}, \bibinfo{person}{Lucian Cojocar}, \bibinfo{person}{Cristiano Giuffrida}, {and} \bibinfo{person}{Herbert Bos}.} \bibinfo{year}{2017}\natexlab{}.
\newblock \showarticletitle{{VUzzer}: Application-Aware Evolutionary Fuzzing}. In \bibinfo{booktitle}{\emph{Proceedings of the 24th Network and Distributed System Security Symposium}} \emph{(\bibinfo{series}{NDSS~'17})}.
\newblock


\bibitem[Schulte et~al\mbox{.}(2022)]%
        {BinaryRewriters}
\bibfield{author}{\bibinfo{person}{Eric Schulte}, \bibinfo{person}{Michael~D. Brown}, {and} \bibinfo{person}{Vlad Folts}.} \bibinfo{year}{2022}\natexlab{}.
\newblock \showarticletitle{A Broad Comparative Evaluation of x86-64 Binary Rewriters}. In \bibinfo{booktitle}{\emph{Proceedings of the 15th Workshop on Cyber Security Experimentation and Test}} \emph{(\bibinfo{series}{CSET~'22})}.
\newblock


\bibitem[Schulte et~al\mbox{.}(2019)]%
        {GTIRB}
\bibfield{author}{\bibinfo{person}{Eric Schulte}, \bibinfo{person}{Jonathan Dorn}, \bibinfo{person}{Antonio Flores-Montoya}, \bibinfo{person}{Aaron Ballman}, {and} \bibinfo{person}{Tom Johnson}.} \bibinfo{year}{2019}\natexlab{}.
\newblock \showarticletitle{{GTIRB}: Intermediate Representation for Binaries}.
\newblock \bibinfo{journal}{\emph{arXiv preprint arXiv:1907.02859}} (\bibinfo{year}{2019}).
\newblock


\bibitem[Serebryany et~al\mbox{.}(2012)]%
        {Asan}
\bibfield{author}{\bibinfo{person}{Konstantin Serebryany}, \bibinfo{person}{Derek Bruening}, \bibinfo{person}{Alexander Potapenko}, {and} \bibinfo{person}{Dmitriy Vyukov}.} \bibinfo{year}{2012}\natexlab{}.
\newblock \showarticletitle{{AddressSanitizer}: A Fast Address Sanity Checker}. In \bibinfo{booktitle}{\emph{Proceedings of the 2012 {USENIX} Annual Technical Conference}} \emph{(\bibinfo{series}{USENIX ATC~'12})}.
\newblock


\bibitem[Shoshitaishvili et~al\mbox{.}(2016)]%
        {angr}
\bibfield{author}{\bibinfo{person}{Yan Shoshitaishvili}, \bibinfo{person}{Ruoyu Wang}, \bibinfo{person}{Christopher Salls}, \bibinfo{person}{Nick Stephens}, \bibinfo{person}{Mario Polino}, \bibinfo{person}{Andrew Dutcher}, \bibinfo{person}{John Grosen}, \bibinfo{person}{Siji Feng}, \bibinfo{person}{Christophe Hauser}, \bibinfo{person}{Christopher Kruegel}, {and} \bibinfo{person}{Giovanni Vigna}.} \bibinfo{year}{2016}\natexlab{}.
\newblock \showarticletitle{{SOK}: (State of) The Art of War: Offensive Techniques in Binary Analysis}. In \bibinfo{booktitle}{\emph{Proceedings of the 2016 {IEEE} Symposium on Security and Privacy}} \emph{(\bibinfo{series}{IEEE S\&P~'16})}.
\newblock


\bibitem[Suh et~al\mbox{.}(2004)]%
        {DIFT}
\bibfield{author}{\bibinfo{person}{G~Edward Suh}, \bibinfo{person}{Jae~W Lee}, \bibinfo{person}{David Zhang}, {and} \bibinfo{person}{Srinivas Devadas}.} \bibinfo{year}{2004}\natexlab{}.
\newblock \showarticletitle{Secure Program Execution via Dynamic Information Flow Tracking}. In \bibinfo{booktitle}{\emph{Proceedings of the 11th International Conference on Architectural Support for Programming Languages and Operating Systems}} \emph{(\bibinfo{series}{ASPLOS~'04})}.
\newblock


\bibitem[van Schaik et~al\mbox{.}(2019)]%
        {RIDL}
\bibfield{author}{\bibinfo{person}{Stephan van Schaik}, \bibinfo{person}{Alyssa Milburn}, \bibinfo{person}{Sebastian Österlund}, \bibinfo{person}{Pietro Frigo}, \bibinfo{person}{Giorgi Maisuradze}, \bibinfo{person}{Kaveh Razavi}, \bibinfo{person}{Herbert Bos}, {and} \bibinfo{person}{Cristiano Giuffrida}.} \bibinfo{year}{2019}\natexlab{}.
\newblock \showarticletitle{{RIDL}: Rogue In-flight Data Load}. In \bibinfo{booktitle}{\emph{Proceedings of the 40th {IEEE} Symposium on Security and Privacy}} \emph{(\bibinfo{series}{IEEE S\&P~'19})}.
\newblock


\bibitem[Wang et~al\mbox{.}(2020)]%
        {KLEESpectre}
\bibfield{author}{\bibinfo{person}{Guanhua Wang}, \bibinfo{person}{Sudipta Chattopadhyay}, \bibinfo{person}{Arnab~Kumar Biswas}, \bibinfo{person}{Tulika Mitra}, {and} \bibinfo{person}{Abhik Roychoudhury}.} \bibinfo{year}{2020}\natexlab{}.
\newblock \showarticletitle{{KLEESpectre}: Detecting Information Leakage through Speculative Cache Attacks via Symbolic Execution}.
\newblock \bibinfo{journal}{\emph{{ACM} Transactions on Software Engineering and Methodology ({TOSEM})}} \bibinfo{volume}{29}, \bibinfo{number}{3} (\bibinfo{year}{2020}), \bibinfo{pages}{1--31}.
\newblock


\bibitem[Wang et~al\mbox{.}(2019)]%
        {oo7}
\bibfield{author}{\bibinfo{person}{Guanhua Wang}, \bibinfo{person}{Sudipta Chattopadhyay}, \bibinfo{person}{Ivan Gotovchits}, \bibinfo{person}{Tulika Mitra}, {and} \bibinfo{person}{Abhik Roychoudhury}.} \bibinfo{year}{2019}\natexlab{}.
\newblock \showarticletitle{{oo7}: Low-Overhead Defense Against {Spectre} Attacks via Program Analysis}.
\newblock \bibinfo{journal}{\emph{{IEEE} Transactions on Software Engineering}} \bibinfo{volume}{47}, \bibinfo{number}{11} (\bibinfo{year}{2019}), \bibinfo{pages}{2504--2519}.
\newblock


\bibitem[Wang et~al\mbox{.}(2016)]%
        {Uroboros}
\bibfield{author}{\bibinfo{person}{Shuai Wang}, \bibinfo{person}{Pei Wang}, {and} \bibinfo{person}{Dinghao Wu}.} \bibinfo{year}{2016}\natexlab{}.
\newblock \showarticletitle{{UROBOROS}: Instrumenting Stripped Binaries with Static Reassembling}. In \bibinfo{booktitle}{\emph{Proceedings of the 23rd {IEEE} International Conference on Software Analysis, Evolution and Reengineering}} \emph{(\bibinfo{series}{SANER~'16})}.
\newblock


\bibitem[Yadavalli and Smith(2019)]%
        {mctoll}
\bibfield{author}{\bibinfo{person}{S.~Bharadwaj Yadavalli} {and} \bibinfo{person}{Aaron Smith}.} \bibinfo{year}{2019}\natexlab{}.
\newblock \showarticletitle{Raising Binaries to {LLVM} {IR} with {MCTOLL} ({WIP} Paper)}. In \bibinfo{booktitle}{\emph{Proceedings of the 20th {ACM} {SIGPLAN/SIGBED} International Conference on Languages, Compilers, and Tools for Embedded Systems}} \emph{(\bibinfo{series}{LCTES~'19})}.
\newblock


\bibitem[Zaitsev(2021)]%
        {jsmn}
\bibfield{author}{\bibinfo{person}{Serge Zaitsev}.} \bibinfo{year}{2021}\natexlab{}.
\newblock \bibinfo{title}{jsmn: Jsmn is a World Fastest {JSON} Parser/Tokenizer}.
\newblock
\newblock
\urldef\tempurl%
\url{https://github.com/zserge/jsmn}
\showURL{%
Retrieved 2023-11-29 from \tempurl}


\bibitem[Zhang and Makris(2020)]%
        {HardwareRealTimeML}
\bibfield{author}{\bibinfo{person}{Yunjie Zhang} {and} \bibinfo{person}{Yiorgos Makris}.} \bibinfo{year}{2020}\natexlab{}.
\newblock \showarticletitle{Hardware-Based Detection of {Spectre} Attacks: A Machine Learning Approach}. In \bibinfo{booktitle}{\emph{Proceedings of the 2020 {Asian} Hardware Oriented Security and Trust Symposium}} \emph{(\bibinfo{series}{AsianHOST~'20})}.
\newblock


\end{thebibliography}
